\def\paperversion{normal}

\def\paperversiondraft{draft}
\def\paperversionnormal{normal}

\ifx\paperversion\paperversionnormal
\else
  \def\paperversion{draft}
\fi

\documentclass[sigplan,screen,nonacm]{acmart}
\def\acmversionanonymous{anonymous}
\def\acmversionjournal{journal}

\makeatletter
\if@ACM@anonymous
  \def\acmversion{anonymous}
\else
  \def\acmversion{journal}
\fi
\makeatother

\usepackage{colortbl}

\usepackage{environ}
\ifx\paperversion\paperversiondraft
\newenvironment{draftonly}{}{}
\else
\NewEnviron{draftonly}{}
\fi

\usepackage[utf8]{inputenc}
\usepackage[T1]{fontenc}
\usepackage{microtype}

\usepackage{overpic}
\usepackage{caption}

\usepackage{xargs}
\usepackage{xparse}
\usepackage{xifthen, xstring}
\usepackage{xspace}
\usepackage{marginnote}
\usepackage{etoolbox}
\usepackage[acronym,shortcuts]{glossaries}
\glsdisablehyper
\usepackage{amsmath}
\usepackage{algorithm}
\usepackage[noend]{algpseudocode}
\usepackage{hyphenat}
\usepackage[shortcuts]{extdash}
\usepackage{subcaption}
\usepackage{tikz}
\usetikzlibrary{arrows.meta}
\usetikzlibrary{calc}
\newlength\capwidth
\usepackage{wrapfig}
\usepackage{multirow}
\usepackage{stfloats}

\makeatletter
\patchcmd{\@addmarginpar}{\ifodd\c@page}{\ifodd\c@page\@tempcnta\m@ne}{}{}
\makeatother
\ifx\paperversion\paperversiondraft
  \makeatletter
  \if@ACM@journal
    \paperwidth=\dimexpr \paperwidth + 3.5cm\relax
    \oddsidemargin=\dimexpr\oddsidemargin + 0cm\relax
    \evensidemargin=\dimexpr\evensidemargin + 0cm\relax
    \marginparwidth=\dimexpr \marginparwidth + 3cm\relax
    \makeatletter
    \long\def\@mn@@@marginnote[#1]#2[#3]{%
      \begingroup
        \ifmmode\mn@strut\let\@tempa\mn@vadjust\else
          \if@inlabel\leavevmode\fi
          \ifhmode\mn@strut\let\@tempa\mn@vadjust\else\let\@tempa\mn@vlap\fi
        \fi
        \@tempa{%
          \vbox to\z@{%
            \vss
            \@mn@margintest
            \if@reversemargin\if@tempswa
                \@tempswafalse
              \else
                \@tempswatrue
            \fi\fi
              \rlap{%
                \if@mn@verbose
                  \PackageInfo{marginnote}{xpos seems to be \@mn@currxpos}%
                \fi
                \begingroup
                  \ifx\@mn@currxpos\relax\else\ifx\@mn@currxpos\@empty\else
                      \kern-\dimexpr\@mn@currxpos\relax
                  \fi\fi
                  \ifx\@mn@currpage\relax
                    \let\@mn@currpage\@ne
                  \fi
                  \if@twoside\ifodd\@mn@currpage\relax
                      \kern\oddsidemargin
                    \else
                      \kern\evensidemargin
                    \fi
                  \else
                    \kern\oddsidemargin
                  \fi
                  \kern 1in
                \endgroup
                \kern\marginnotetextwidth\kern\marginparsep
                \vbox to\z@{\kern\marginnotevadjust\kern #3
                  \vbox to\z@{%
                    \hsize\marginparwidth
                    \linewidth\hsize
                    \kern-\parskip
                    \marginfont\raggedrightmarginnote\strut\hspace{\z@}%
                    \ignorespaces#2\endgraf
                    \vss}%
                  \vss}%
              }%
          }%
        }%
      \endgroup
    }
    \makeatother
  \else
    \paperwidth=\dimexpr \paperwidth + 6cm\relax
    \oddsidemargin=\dimexpr\oddsidemargin + 3cm\relax
    \evensidemargin=\dimexpr\evensidemargin + 3cm\relax
    \marginparwidth=\dimexpr \marginparwidth + 3cm\relax
  \fi
  \makeatother
\fi

\usepackage[textsize=tiny]{todonotes}
\usepackage[normalem]{ulem}

\makeatletter
\font\uwavefont=lasyb10 scaled 652
\newcommand\colorwave[1][blue]{\bgroup\markoverwith{\lower3\p@\hbox{\uwavefont\textcolor{#1}{\char58}}}\ULon}

\newcommand\highlight[2]{{\color{#1}{\colorwave[#1]{#2}}}}
\makeatother

\makeatletter
\newcommand\InFloat[2]{\ifnum\@floatpenalty<0\relax#1\else#2\fi}
\makeatother

\newboolean{inComment}
\setboolean{inComment}{false}

\ifx\paperversion\paperversiondraft
\newcommand\createtodoauthor[2]{
  \def\tmpdefault{emptystring}
  \expandafter\newcommand\csname #1\endcsname[2][\tmpdefault]{
    \ifthenelse{\boolean{inComment}}{
      \PackageError{paper-template}{Comments in comments not supported}{}
    }{}\setboolean{inComment}{true}
    \def\tmp{##1}
    \InFloat{
        \smash{
	  \marginnote{
	    \todo[inline,linecolor=#2,backgroundcolor=#2,bordercolor=#2]
	      {\textbf{#1 (Figure):} ##2}
          }
        }
    }{\ifthenelse{\equal{\tmp}{\tmpdefault}} 
      {\todo[linecolor=#2,backgroundcolor=#2,bordercolor=#2]{\textbf{#1:} ##2}\ignorespaces}
      {\ifthenelse{\equal{##2}{}} 
        {\highlight{#2}{##1}}
        {\highlight{#2}{##1}\todo[linecolor=#2,backgroundcolor=#2,bordercolor=#2]
	  {\textbf{#1:} ##2}
	}
      }
    }
    \setboolean{inComment}{false}
  }
}
\else
\newcommand\createtodoauthor[2]{%
\expandafter\newcommand\csname #1\endcsname[2][]{##1}%
}%
\fi

\usepackage[frozencache,cachedir=minted-cache]{minted}
\usepackage{etoolbox}

\makeatletter
\ifcsdef{minted@optlistcl@quote}
{
\ifwindows
  \renewcommand{\minted@optlistcl@quote}[2]{%
    \ifstrempty{#2}{\detokenize{#1}}{\detokenize{#1="#2"}}}
\else
  \renewcommand{\minted@optlistcl@quote}[2]{%
    \ifstrempty{#2}{\detokenize{#1}}{\detokenize{#1='#2'}}}
\fi

\newcommand{\minted@def@optcl@novalue}[2]{%
  \define@booleankey{minted@opt@g}{#1}%
    {\minted@addto@optlistcl{\minted@optlistcl@g}{#2}{}%
     \@namedef{minted@opt@g:#1}{true}}
    {\@namedef{minted@opt@g:#1}{false}}
  \define@booleankey{minted@opt@g@i}{#1}%
    {\minted@addto@optlistcl{\minted@optlistcl@g@i}{#2}{}%
     \@namedef{minted@opt@g@i:#1}{true}}
    {\@namedef{minted@opt@g@i:#1}{false}}
  \define@booleankey{minted@opt@lang}{#1}%
    {\minted@addto@optlistcl@lang{minted@optlistcl@lang\minted@lang}{#2}{}%
     \@namedef{minted@opt@lang\minted@lang:#1}{true}}
    {\@namedef{minted@opt@lang\minted@lang:#1}{false}}
  \define@booleankey{minted@opt@lang@i}{#1}%
    {\minted@addto@optlistcl@lang{minted@optlistcl@lang\minted@lang @i}{#2}{}%
     \@namedef{minted@opt@lang\minted@lang @i:#1}{true}}
    {\@namedef{minted@opt@lang\minted@lang @i:#1}{false}}
  \define@booleankey{minted@opt@cmd}{#1}%
      {\minted@addto@optlistcl{\minted@optlistcl@cmd}{#2}{}%
        \@namedef{minted@opt@cmd:#1}{true}}
      {\@namedef{minted@opt@cmd:#1}{false}}
}
\minted@def@optcl@novalue{customlexer}{-x}
}
{
}
\makeatother

\usepackage{tikz}
\usetikzlibrary{arrows}
\usetikzlibrary{shapes}

\tikzset{
  circledstyle/.style={
    shape=circle,
    #1,
    font=\tt\small,
    inner sep=0pt,
    outer sep=0pt,
    minimum size=1.2em,
    text=black
  }
}

\DeclareRobustCommand{\circledbase}[3][]{%
    \tikz[baseline=(char.base)]{\node[circledstyle, fill=#2] (char) {#3\strut};}%
}

\usepackage[norefs,nocites]{refcheck}

\makeatletter
\newcommand{\refcheckize}[1]{%
  \expandafter\let\csname @@\string#1\endcsname#1%
  \expandafter\DeclareRobustCommand\csname relax\string#1\endcsname[1]{%
    \csname @@\string#1\endcsname{##1}\wrtusdrf{##1}}%
  \expandafter\let\expandafter#1\csname relax\string#1\endcsname
}
\makeatother

\refcheckize{\autoref}

\ifx\acmversion\acmversionjournal
  \acmJournal{PACMPL}
  \acmVolume{1}
  \acmNumber{1}
  \acmArticle{1}
  \acmYear{2017}
  \acmMonth{1}
  \acmDOI{10.1145/nnnnnnn.nnnnnnn}
  \startPage{1}
\else
  \acmConference[PL'17]{ACM SIGPLAN Conference on Programming Languages}{January 01--03, 2017}{New York, NY, USA}
  \acmYear{2017}
  \acmISBN{978-x-xxxx-xxxx-x/YY/MM}
  \acmDOI{10.1145/nnnnnnn.nnnnnnn}
  \startPage{1}
\fi

\usepackage{relsize}
\setminted{fontsize=\relsize{-1}}

\makeatletter
\ifcsdef{MintedExecutable}
{
  \newminted[mlir]{tools/lexers/MLIRLexer.py:MLIRLexerOnlyOps}{mathescape, breaklines, escapeinside=||}
  \newminted[xdsl]{tools/lexers/MLIRLexer.py:MLIRLexer}{mathescape, breaklines, escapeinside=||, style=murphy}
  \newminted[lean4]{tools/lexers/Lean4Lexer.py:Lean4Lexer}{mathescape}
}
{
  \ifcsdef{minted@optlistcl@quote}
  {
    \newminted[mlir]{tools/lexers/MLIRLexer.py:MLIRLexerOnlyOps}{customlexer, mathescape, breaklines, escapeinside=||}
    \newminted[xdsl]{tools/lexers/MLIRLexer.py:MLIRLexer}{customlexer, mathescape, breaklines, escapeinside=||, style=murphy}
    \newminted[lean4]{tools/lexers/Lean4Lexer.py:Lean4Lexer}{customlexer, mathescape}
  }
  {
    \newminted[mlir]{tools/lexers/MLIRLexer.py:MLIRLexerOnlyOps -x}{mathescape, breaklines, escapeinside=||}
    \newminted[xdsl]{tools/lexers/MLIRLexer.py:MLIRLexer -x}{mathescape, breaklines, escapeinside=||, style=murphy}
    \newminted[lean4]{tools/lexers/Lean4Lexer.py:Lean4Lexer -x}{mathescape}
  }
}
\makeatother

\definecolor{darkGray}{HTML}{353535}
\definecolor{pairedNegOneLightGray}{HTML}{cacaca}
\definecolor{pairedNegTwoDarkGray}{HTML}{827b7b}
\definecolor{pairedOneLightBlue}{HTML}{a6cee3}
\definecolor{pairedTwoDarkBlue}{HTML}{1f78b4}
\definecolor{pairedThreeLightGreen}{HTML}{b2df8a}
\definecolor{pairedFourDarkGreen}{HTML}{33a02c}
\definecolor{pairedFiveLightRed}{HTML}{fb9a99}
\definecolor{pairedSixDarkRed}{HTML}{e31a1c}
\definecolor{classBrown}{HTML}{a06227}

\definecolor{mygray}{gray}{0.6}

\newcommand{\gray}[1]{\textcolor{mygray}{#1}}
\usepackage{xfp}
\makeatletter
\AtBeginDocument{%
  \begingroup
  \normalsize
  \let\tmp@n@s\f@size
  \let\tmp@n@b\f@baselineskip
  \small
  \let\tmp@s@s\f@size
  \let\tmp@s@b\f@baselineskip
  \footnotesize
  \let\tmp@f@s\f@size
  \let\tmp@f@b\f@baselineskip
  \xdef\semismall@size{\fpeval{(\tmp@n@s+\tmp@s@s)/2.07}}%
  \xdef\semismall@baselineskip{\fpeval{(\tmp@n@b+\tmp@n@b)/2.07}}%
  \xdef\semifootnotesmall@size{\fpeval{(\tmp@f@s+\tmp@f@s)/1.75}}%
  \xdef\semifootnotesmall@baselineskip{\fpeval{(\tmp@f@b+\tmp@f@b)/1.75}}%
  \xdef\semicaptionsmall@size{\fpeval{(\tmp@n@s+\tmp@s@s)/2.25}}%
  \xdef\semicaptionsmall@baselineskip{\fpeval{(\tmp@n@b+\tmp@n@b)/2.25}}%
  \endgroup
}
\newcommand{\semismall}{\fontsize{\semismall@size}{\semismall@baselineskip}\selectfont}
\newcommand{\semifootnotesmall}{\fontsize{\semifootnotesmall@size}{\semifootnotesmall@baselineskip}\selectfont}
\newcommand{\mlirdialect}[1]{%
  \texttt{\semismall\textcolor{darkGray}{#1}}%
}

\newcommand{\mlirop}[2]{%
  \texttt{\semismall\textcolor{darkGray}{#1.#2}}%
}

\newcommand{\code}[1]{{\semismall{\texttt{#1}}}}

\createtodoauthor{grosser}{pairedOneLightBlue}
\createtodoauthor{sasha}{pairedTwoDarkBlue}
\createtodoauthor{jules}{pairedThreeLightGreen}
\createtodoauthor{sam}{pairedFourDarkGreen}
\createtodoauthor{jianyi}{pairedFiveLightRed}
\createtodoauthor{authorSix}{pairedSixDarkRed}
\createtodoauthor{bjorn}{pairedNegOneLightGray}

\newacronym{ir}{IR}{intermediate representation}
\newacronym{ssa}{SSA}{static single assignment}
\newacronym{fp}{FP}{floating\-/point}
\newacronym{mlir}{MLIR}{Multi-Level Intermediate Representation}
\newacronym{cse}{CSE}{common sub-expression elimination}
\newacronym{aegraph}{ægraph}{acyclic e\-/graph}
\newacronym{cfg}{CFG}{control flow graph}
\newacronym{ilp}{ILP}{Integer Linear Programming}
\newacronym{pdl}{PDL}{Pattern Definition Language}
\newacronym{ulp}{ULP}{Units in the Last Place}
\newacronym{hdl}{HDL}{Hardware Description Language}
\newacronym{es}{EqSat}{Equality saturation}

\graphicspath{{./images/}}

\makeatletter
\newcommand\requiredelimiter[2][########]{%
  \ifdefined#2%
    \def\@temp{\def#2#1}%
    \expandafter\@temp\expandafter{#2}%
  \else
    \@latex@error{\noexpand#2undefined}\@ehc
  \fi
}
\@onlypreamble\requiredelimiter
\makeatother

\newcommand\newdelimitedcommand[2]{
\expandafter\newcommand\csname #1\endcsname{#2}
\expandafter\requiredelimiter
\csname #1 \endcsname
}

\newdelimitedcommand{toolname}{Tool}

\makeatletter
\newcommand{\cunderline}[3][1.5pt]{%
  \def\cul@thick{#1}\def\cul@color{#2}%
  \mathpalette\cul@draw{#3}%
}
\newcommand{\cul@draw}[2]{
  \sbox\z@{$\m@th#1#2$}%
  \hbox{%
    \copy\z@
    \kern-\wd\z@
    \lower3.7\p@\hbox{\color{\cul@color}%
      \vrule width\wd\z@ height\cul@thick depth\z@}%
  }%
}
\makeatother

\usepackage{booktabs}
\newcommand{\ra}[1]{\renewcommand{\arraystretch}{#1}}

\usepackage[verbose]{newunicodechar}
\newunicodechar{₁}{\ensuremath{_1}}
\newunicodechar{₂}{\ensuremath{_2}}
\newunicodechar{∀}{\ensuremath{\forall}}
\newunicodechar{α}{\ensuremath{\alpha}}
\newunicodechar{β}{\ensuremath{\beta}}

\DeclareRobustCommand{\circled}[2][]{%
    \ifthenelse{\isempty{#1}}%
        {\circledbase{pairedOneLightBlue}{#2}}%
        {\autoref{#1}: \hyperref[#1]{\circledbase{pairedOneLightBlue}{#2}}}%
}

\DeclareRobustCommand{\circledgreen}[2][]{%
    \ifthenelse{\isempty{#1}}%
        {\circledbase{pairedThreeLightGreen}{#2}}%
        {\autoref{#1}: \hyperref[#1]{\circledbase{pairedThreeLightGreen}{#2}}}%
}

\begin{document}

\title[]{E-Graphs as a Persistent Compiler Abstraction}

\author{Jules Merckx}
\affiliation{
  \institution{Ghent University}
  \city{Ghent}
  \country{Belgium}
}
\email{jules.merckx@ugent.be}

\author{Alexandre Lopoukhine}
\affiliation{
  \institution{University of Cambridge}
  \city{Cambridge}
  \country{United Kingdom}
}
\email{sasha.lopoukhine@cl.cam.ac.uk}

\author{Samuel Coward}
\affiliation{
  \institution{University College London}
  \city{London}
  \country{United Kingdom}
}
\email{sam.coward@ucl.ac.uk}

\author{Bjorn De Sutter}
\affiliation{
  \institution{Ghent University}
  \city{Ghent}
  \country{Belgium}
}
\email{bjorn.desutter@ugent.be}

\author{Jianyi Cheng}
\affiliation{
  \institution{University of Edinburgh}
  \city{Edinburgh}
  \country{United Kingdom}
}
\email{jianyi.cheng@ed.ac.uk}

\author{Tobias Grosser}
\affiliation{
  \institution{University of Cambridge}
  \city{Cambridge}
  \country{United Kingdom}
}
\email{tobias.grosser@cst.cam.ac.uk}

\authorsaddresses{}

\begin{abstract}

  Recent algorithmic advances have made equality saturation an appealing technique for program optimization, avoiding the phase-ordering problem by separating the discovery of equivalent expressions from optimal expression selection.
  Existing work leveraging equality saturation in compilers uses either external equality saturation libraries or custom implementations that are coupled to the specific application.
  These approaches are inherently limited, as the first discards semantic equivalences when translating back from the external library, while the second restricts equality saturation to a single level of abstraction.
  We propose an alternative approach that represents an e-graph natively in code, facilitating the application of constructive compiler passes that maintain the e-graph state throughout the compilation flow.
  We present \emph{Tamagoyaki}, an implementation of this approach in MLIR and demonstrate its versatility through partial re-implementations of two equality saturation applications.
  Our software case study improves performance by $1.18\times$ on average, whilst our hardware case study reduces circuit delay by up to 11\% over standard equality saturation.\grosser{Can we specific here? Instead of generic "software case study" say 
  that we improve over Herbie and ROVER (and say what they do)"}
  Building on reusable compiler infrastructure, our work expands the scope of equality saturation, making it persistent across and interleavable with other analyses and transformations.


\end{abstract}

\ifx\acmversion\acmversionanonymous
  \renewcommand\footnotetextcopyrightpermission[1]{} 
\fi
\settopmatter{printacmref=false} 

\maketitle

\section{Introduction}

\ac{es}~\cite{Tate2009EqualityOptimization} has been applied to a remarkable breadth of optimization challenges, including
floating-point accuracy~\cite{Panchekha2015Herbie},
hardware datapath optimization~\cite{Coward2024ROVER:Rewriting,Wanna2023MultiplierRewriting},
logic synthesis~\cite{Chen2024E-Syn:Synthesis},
3D~CAD model synthesis~\cite{Nandi2020SynthesizingTransformations},
low-level arithmetic optimization~\cite{cranelift},
HLS super-optimization~\cite{Cheng2024SEER:MLIR},
and high-level language compiler optimization~\cite{Merckx2026EqualitySaturationOptimizing,dialegg2025}.
Its most attractive quality is that it does not suffer from the phase-ordering problem, where one optimization can prevent another from being applied.
\ac{es} tracks all equivalent expressions discovered via rewriting in a single data structure, the e-graph, from which users then extract an optimal representation, potentially unreachable using classical rewriting.

\begin{figure}
  \centering
  \begin{subfigure}{0.32\columnwidth}
    \centering
    \includegraphics{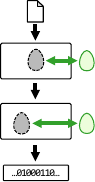}
    \caption{External library}
    \label{fig:compilation_egg}
  \end{subfigure}%
  \begin{subfigure}{0.32\columnwidth}
    \centering
    \includegraphics{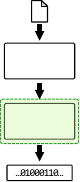}
    \caption{Bespoke pass}
    \label{fig:compilation_cranelift}
  \end{subfigure}%
  \begin{subfigure}{0.32\columnwidth}
    \centering
    \includegraphics{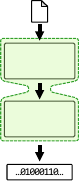}
    \caption{Embedded in IR}
    \label{fig:compilation_equivalence}
  \end{subfigure}
  \caption{Existing \ac{es} approaches either (a) go back and forth between the compiler and an external tool or library such as egg~\cite{Willsey2021Egg}, or (b) hardcode e-graphs within part of the compiler for a specialized use case~\cite{cranelift}.  
  \emph{Tamagoyaki} (c) embeds e-graphs natively into a compiler IR such that equalities now persist across abstraction levels.}
  \label{fig:compilation}
\end{figure}

Modern compilers work as a sequence of transformations and analysis passes, often at multiple levels of abstraction~\cite{dragonbook}.
Existing work applying \ac{es} to code optimization typically converts a program at a particular abstraction level into an e-graph, applies \ac{es}, then returns a single optimized representation for the next pass (\autoref{fig:compilation_egg})~\cite{dialegg2025,Cheng2024SEER:MLIR}.
This imposes a potentially lossy translation overhead, discards equivalences, and prevents interleaving \ac{es} with other passes such as abstraction lowering, inlining, or other transformations that cannot take part in \ac{es}.
A notable embedding of \ac{es} natively in a compiler flow is Cranelift~\cite{cranelift}. It embeds a custom e-graph-inspired data structure to represent its \ac{ir}, optimizing at the arithmetic level (\autoref{fig:compilation_cranelift}).
However, it is neither a generic \ac{es} framework nor a generic compiler framework, limiting its use.

We present an alternative approach that embeds e-graph primitives in an extensible compiler \ac{ir} (\autoref{fig:compilation_equivalence}).
With the addition of an e-graph rewriting procedure that reinterprets MLIR's~\cite{mlir2021} pattern matcher based on the \ac{pdl}~\cite{pdl}, we implement \ac{es} within the compiler itself.
This obviates the need to reimplement \ac{es} support again and again for each domain-specific \acs{ir}.\bjorn{I added this sentence in preparation of the motivation of better scaling.}
It retains the expressibility of general \ac{es} libraries, but eliminates translation overhead.
Furthermore, the compiler can keep track of equivalent expressions at different levels of abstraction, and exploit them throughout compilation. 

\pagebreak
Our contributions are:
\begin{itemize}
  \item novel primitives for encoding e-graphs in \ac{ir}, letting equalities persist throughout compilation (\autoref{subsec:equivalence_dialect}),
\item a reinterpretation of e-graph invariants in SSA, obviating an external union-find structure (\autoref{subsubsec:invariants}),
\item compiler-native e-matching using existing declarative rewriting infrastructure (\autoref{subsubsec:ematching}),
\item decoupling e-graph extraction into separate selection and replacement passes over plain \ac{ir}, allowing chaining of heterogeneous cost models and e-graph pruning with standard IR manipulation (\autoref{subsubsec:extraction}),
\item bidirectional reuse between existing dataflow analyses and e-class analyses (\autoref{subsec:eclass_analyses}),
\item \emph{Tamagoyaki}, our implementation adding cross-dialect \ac{es} support to the extensible MLIR framework, 
\item two case study demos of its potential for both software (\autoref{sec:evaluation_herbie}) and hardware (\autoref{sec:evaluation_rover}) optimization.
\end{itemize}


\section{Motivating Persistent E-Graphs}\label{sec:motivating}
Embedding an e-graph in a multi-level \ac{ir} can unlock new optimization capabilities by exploiting persistent equalities.
Consider a simple combinational circuit to implement
\[
\code{out = (a << m) * (b << n)},
\]
where \code{a} and \code{b} are 32-bit inputs, \code{m} and \code{n} are 5-bit shift amounts, and \code{out} is a 64-bit output. All the operators operate at 64 bits.
CIRCT, an MLIR-based hardware compiler framework, uses \emph{IR dialects} to model domain-specific constructs in IR~\cite{Eldridge2021MLIRInfrastructure}. 
In CIRCT, we get a \mlirdialect{comb} dialect representation that explicitly zero extends operands to 64-bits:
\begin{mlir}
    |\textcolor{pairedNegOneLightGray}{\%a\_i64 = comb.concat \%c0\_i32, \%a : i32, i32}|
    |\textcolor{pairedNegOneLightGray}{\%b\_i64 = comb.concat \%c0\_i32, \%b : i32, i32}|
\end{mlir}

\begin{figure}[b]
    \centering
    \includegraphics[]{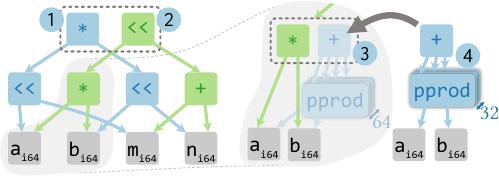}

    \vspace{-0.5cm}   
    \begin{minipage}[t]{0.45\columnwidth}
        \centering
        \subcaption{Single-level}
        \label{fig:shiftmul-comb}
    \end{minipage}
    \hfill
    \begin{minipage}[t]{0.25\columnwidth}
        \centering
        \subcaption{Multi-level}
        \label{fig:shiftmul-datapath}
    \end{minipage}
    \hfill
    \begin{minipage}[t]{0.25\columnwidth}
        \centering
        \subcaption{Destructive}
        \label{fig:shiftmul-destructive}
    \end{minipage}

    \caption{
        (a) Applying only \mlirdialect{comb} level rewrites to the initial design \circled{1}, Tamagoyaki discovers alternative \circled{2}. 
        (b) Applying \mlirdialect{comb} to \mlirdialect{datapath} lowering during \ac{es}, Tamagoyaki discovers an alternative representation \circled{3}.
        (c) Applying a destructive \code{canonicalize} pass simplifies the alternative by reducing the number of partial products \circled{4}.
    }
    \label{fig:shift_mul}
\end{figure}

Unlike prior work that used external frameworks~\cite{Coward2022Rover,dialegg2025}, Tamagoyaki lets us replicate the exploration of equivalent expressions (as modeled in an e-graph, \autoref{subsec:background:equivalence}) directly in the \ac{ir}.
Developers thus need not implement (potentially lossy) dialect-specific conversions to and from such external frameworks like egg~\cite{Willsey2021Egg}.
The IR we start from \emph{already} forms an e-graph (\circled[fig:shiftmul-comb]{1}), albeit without any equivalences between values yet, i.e., without any alternative implementations of the circuit. 

Applying rewrites from existing work~\cite{Coward2022Rover} to this CIRCT IR, Tamagoyaki adds the following promising alternative option (\circled[fig:shiftmul-comb]{2}):
\begin{mlir}
\end{mlir}
When we synthesize both options, the second is 11\% faster, as a result of CIRCT's further compilation simplifying its multiplication because its operands are zero extended, unlike those in the original IR.
We hence want the compiler to select the second option for further compilation and synthesis.

To select the option with the smallest latency, the typical egg approach uses a simple local cost model. In this case, such a model will determine an operator's latency based on its bitwidth~\cite{Coward2022Rover,kong2026improving}. However, since both options appear to have the same bitwith, the typical egg-like approach and model cannot differentiate between them.

The core problem is the cost model's poor prediction of downstream optimizations.
Tamagoyaki can do better, however, because its persistent IR-embedded e-graph lets us apply downstream optimization passes directly to the e-graph before selecting the best option.
This lets our cost model ``see'' the consequences of downstream optimizations. 

First, we express the lowering from \mlirdialect{comb} to the lower-level \mlirdialect{datapath} dialect as rewrites that we apply to the e-graph.  
\begin{mlir}
\end{mlir}  
\vspace{-12pt}  
\begin{center}
\begin{tikzpicture}
\draw[-{Triangle[length=5pt,width=8pt]}, pairedNegOneLightGray, line width=4pt] (0,0.4) -- (0,0);
\end{tikzpicture}
\end{center}
\vspace{-13pt}  
\begin{mlir}
\end{mlir}
The \code{partial\_product} operation decomposes multiplication into a list of shifted bitvectors whose sum equals the full product~\cite{deDinechin2024Application-SpecificArithmetic}. 
In Tamagoyaki, these rewrites add another option to the e-graph (\circled[fig:shift_mul]{3}), i.e., to the IR. This option enables multi-level exploration. 

Such exploration across multiple hardware abstraction layers with \ac{es} has already shown to be beneficial in existing work~\cite{kong2026improving}. However, they relied on an external library, whilst Tamagoyaki embeds the exploration in the \ac{ir} itself. This lets us combine \ac{es} with existing compiler passes on MLIR dialects, 
without developers having to change any passes.
In this case, the existing CIRCT \code{canonicalize} pass can simplify the \code{partial\_product} operator, because only 32 of the 64 partial product bitvectors can be non-zero. 
Applying \code{canonicalize}, Tamagoyaki destructively simplifies the design (\circled[fig:shiftmul-destructive]{4}). 
We could resume rewriting, but if we now extract from this e-graph, extended to model \mlirdialect{datapath} operations, using the cost model we indeed select a design that uses 42\% fewer logic gates whilst being 11\% faster.

Tamagoyaki automates the exploration flow described above.
It treats standard MLIR \ac{ir} as an e-graph.
Using our \mlirdialect{equivalence} dialect, we can add equivalences to the e-graph.
Furthermore, we extend MLIR's rewrite pattern infrastructure to apply rewrites on this IR with equivalences.
With Tamagoyaki, one can apply rewrites across different abstraction levels, and even apply existing optimization passes to the e-graph itself.
Since \mlirdialect{equivalence} is just another dialect, it easily interoperates with existing passes, as we have seen, allowing the e-graph to \emph{persist}.

For now, this example has demonstrated how Tamagoyaki combines CIRCT's multiple abstraction layers with \ac{es} to unlock new optimization capabilities. Later we provide a full case study of our Tamagoyaki-based circuit optimizer (\autoref{sec:evaluation_rover}). An additional software optimization case study (\autoref{sec:evaluation_herbie}) will demonstrate Tamagoyaki's applicability to multiple domains and optimization objectives.

\section{Background}
\label{sec:background}

Tamagoyaki integrates \ac{es} into compiler infrastructure widely adopted in industry.
It is implemented in MLIR~\cite{mlir2021}, a user-extensible compiler framework that leverages \acs{ssa} and \textit{regions}, and extends MLIR's existing rewrite infrastructure.

\subsection{Static Single Assignment with Regions}\label{subsec:background:ssa}

\begin{figure}[t]
  \centering
  \includegraphics[width=0.92\columnwidth]{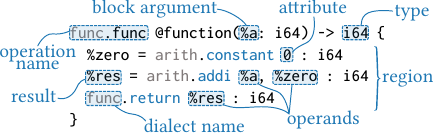}
  \caption{MLIR incorporates user extensibility by design, combined with properties such as \acrshort{ssa}, to provide a productive environment for exploring novel compiler techniques.}
  \label{fig:mlir-primer}
  \vspace{-0.5em}
\end{figure}

\begin{figure}[t]
  \centering
  \includegraphics[width=0.65\columnwidth]{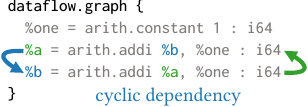}
  \caption{MLIR supports cyclic code in \textit{graph} regions.}
  \label{fig:mlir-graph-region}
\end{figure}

Our work leverages key advances in modern compiler \ac{ir} infrastructure (\autoref{fig:mlir-primer}).
\textit{Operations} like \mlirop{arith}{addi} represent units of computation, with compile-time information represented with \textit{attributes}, and run-time information represented with \textit{values}.
New values can be defined as \textit{results} of operations, e.g.,~\code{\%res}, or as \textit{block} arguments, e.g.,~\code{\%a}.
Values are defined exactly once, a property known as \ac{ssa}~\cite{Lattner2004LLVM,gcc-compiler}, simplifying analyses and transformations during compilation.
Values store pointers to their defining operations or blocks, and to the operations that use them as \textit{operands}, forming an object graph.
In MLIR, operation definitions are grouped into \textit{dialects} like \mlirdialect{arith} and \mlirdialect{func}, which serve as name\-spaces for related definitions.

Operations in MLIR can contain regions, which contain blocks, which themselves contain operations, forming a recursive \ac{ir} structure.
Operation definitions restrict the regions they contain to be one of two kinds.
The most common is a region comprising a \ac{cfg}, which may have multiple blocks, whose successors must all be directly contained within the same region, with the key restriction that value definitions must precede their uses.
For example, the \mlirop{func}{func} operation represents a function definition, and restricts its function body to be a \ac{cfg} region (\autoref{fig:mlir-primer}).
To encode cyclic code, operations may choose instead to contain \textit{graph} regions, which must have a single block, but allow uses of values to precede their definitions, making them useful to represent, for example, dataflow graphs (\autoref{fig:mlir-graph-region}).

\subsection{Declarative IR Rewrite Patterns}\label{subsec:background:pdl}

We extend \mlirdialect{pdl}, MLIR's declarative \ac{ir} pattern rewrite infrastructure used in projects such as IREE~\cite{TinyIREE} to optimize code.

\begin{listing}
  
\semismall
\begin{minted}[escapeinside=||]{text}
pdl.pattern : benefit(1) {
  %0 = pdl.type |\textcolor{pairedNegTwoDarkGray}{// Arbitrary type to match}|
  %a = pdl.operand |\textcolor{pairedNegTwoDarkGray}{// Free variable to match}|
  %1 = pdl.operation |"arith.andi"|( |\textcolor{pairedNegTwoDarkGray}{// Pattern to match: a \& a}|
      %a, %a: !pdl.value, !pdl.value
    ) -> (%0: !pdl.type)
  |\textcolor{pairedNegTwoDarkGray}{// Imperative Rewrite Procedure:}|
  pdl.rewrite %1 { |\textcolor{pairedNegTwoDarkGray}{// Rewrite matched pattern (a \& a) into a}|
    pdl.replace %1 with (%a: !pdl.value)
  }
}
\end{minted}
\semismall{}
\vspace{-10pt}
\begin{center}
\begin{tikzpicture}
\draw[-{Triangle[length=5pt,width=8pt]}, pairedNegOneLightGray, line width=4pt] (0,0.4) -- (0,0);
\end{tikzpicture}
\end{center}

\begin{minted}[escapeinside=||]{text}
pdl_interp.func @matcher(%arg0: !pdl.operation) {
  |\textcolor{pairedNegTwoDarkGray}{// Check matched operation has at least 2 arguments:}|
  %0 = pdl_interp.get_operand 1 of %arg0
  %1 = pdl_interp.get_defining_op of %0 : !pdl.value
  |\textcolor{pairedNegTwoDarkGray}{// if previous check failed then bail out ({\textasciicircum}bb1),}|
  |\textcolor{pairedNegTwoDarkGray}{// else continue ({\textasciicircum}bb2):}|
  pdl_interp.is_not_null %1: !pdl.operation -> ^bb2, ^bb1
^bb1: |\textcolor{pairedNegTwoDarkGray}{// denotes a basic block}|
  pdl_interp.finalize |\textcolor{pairedNegTwoDarkGray}{// bail out (no match)}|
^bb2: |\textcolor{pairedNegTwoDarkGray}{// Check matched operation is \&:}|
  pdl_interp.check_operation_name of %arg0 is "arith.andi"
    -> ^bb3, ^bb1
^bb3: |\textcolor{pairedNegTwoDarkGray}{// Check number of operands:}|
  pdl_interp.check_operand_count of %arg0 is 2 -> ^bb4, ^bb1
\end{minted}
\vspace{-5pt}
\begin{center}
$\mathbf{\dots}$
\end{center}

  \caption{Declarative rewrite patterns in MLIR's \mlirdialect{pdl} dialect are lowered to an imperative state machine comprising simple instructions, expressed in the \mlirdialect{pdl\_interp} dialect.}
  \label{list:pdl}
\end{listing}

\paragraph{The Pattern Description Language}
Rewrites defined in \mlirdialect{pdl} are composed of two parts~\cite{pdl}, the first a declarative matching pattern, and the second an imperative rewrite procedure, e.g., \code{a \& a} $\rightarrow$ \code{a}~(\autoref{list:pdl}).
Patterns are defined to be applied destructively, meaning that the matched pattern is usually replaced in the \ac{ir}.
Static types of operation results and operands can be matched to describe complex type constraints.
To address cases where multiple patterns match on the same \ac{ir}, patterns declare a \textit{benefit} value, which specifies the precedence of the rewrites.

\paragraph{Interpreting PDL}
Declarative rewrites in the \mlirdialect{pdl} dialect are lowered into an imperative state machine expressed using \mlirdialect{pdl\_interp} operations~(\autoref{list:pdl}).
This lowering can combine multiple patterns into one search routine, reusing information from different patterns and exiting as soon as none of the patterns can be matched anymore.
Operations in the \mlirdialect{pdl\_interp} dialect encode either atomic matching steps, such as checking an operation name, or primitive control flow, such as checking for the next step in the matching procedure, with each operation corresponding to an interpreter instruction.
Finally, MLIR converts this \ac{ir} into a bytecode format for efficient interpretation.

\subsection{Equivalence Graphs and Equality Saturation}\label{subsec:background:equivalence}

\ac{es} is an established technique that explores the space of equivalent expressions before selecting the optimal one~\cite{Tate2009EqualityOptimization,Willsey2021Egg}.
The equivalence graph (e-graph) data structure provides a compact representation of equivalence classes (e-classes) of terms~\cite{Nelson1980TechniquesVerification}.
\ac{es} is the process of expanding this graph by applying rewrite rules.
After this, a cost model is used to select a desired expression.
This approach avoids the need for a fixed order of rewrite application, avoiding the phase-ordering problem.

E-graphs achieve an efficient representation of equivalent expressions by introducing e-classes.
Expressions are represented with e-nodes composed of a function symbol and a list of children.
Instead of referring to other e-nodes, the children refer to e-classes, which are typically represented with a union-find data structure~\cite{Tarjan1975UnionFind}.
This indirection allows an e-graph with $O(n)$ nodes to represent $O(2^n)$ expressions~\cite{Nelson1980TechniquesVerification}.

Rather than replacing existing terms as is common in traditional compilers, terms are instead added to the data structure.
The introduction of e-classes complicates term rewriting, demanding custom matching procedures, known as e-matching~\cite{DeMoura2007EfficientSolvers,zhang2022relational, Tate2009EqualityOptimization}.
E-matching is responsible for backtracking and considering different candidate e-nodes when matching.
Key to efficient e-matching is exploiting correlated variables to prune candidates early~\cite{DeMoura2007EfficientSolvers}.
For example, consider the following pattern containing two variables:
\[
  f({\color{pairedTwoDarkBlue}x_1}, g({\color{pairedTwoDarkBlue}x_1}), h({\color{pairedFourDarkGreen}x_2}))
\]
Exploiting the multiple occurrences of $x_1$ reduces the number of candidates to be processed: if the first arguments to $f$ and $g$ do not agree during matching, different candidates for $x_2$ should not even be considered.
More generally, multiple different patterns may share subpatterns. This can be exploited to further reduce the candidates to be considered~\cite{DeMoura2007EfficientSolvers}.

To efficiently reason over equalities, several invariants must be maintained in the e-graph, to ensure that no redundant equivalences are introduced.
A recent innovation by the egg library~\cite{Willsey2021Egg} is deferred \emph{rebuilding}, where restoration of one of the invariants is deferred until after rewrites have been applied, thereby avoiding redundant work.

Another innovation by egg is \emph{e-class analysis}.
This extends \ac{es} with the ability to track and propagate analysis information.
These analyses are most commonly used to validate conditional rewrites.
While not central to \ac{es}, e-class analysis is a powerful tool~\cite{Coward2023CombiningInterpretation} that is supported by most modern e-graph frameworks~\cite{Willsey2021Egg, Zhang2023Egglog}.

Once \ac{es} terminates, either once no new matches can be found or a computational limit is reached, the optimal term is found using a cost model, a process known as extraction.
Extraction can be guided by either a cheap local cost model~\cite{Willsey2021Egg} or a more expensive global procedure~\cite{Wang2020SPORES:Algebra, Sun2024EGraphsTreewidth, Goharshady2024FastGraphs,Cai2025}.
The choice of the cost model and extraction procedure is determined by the application.

\typeout{Text width: \the\textwidth}
\typeout{Column width: \the\columnwidth}
\section{A Persistent E-Graph Abstraction}
\label{sec:embedding_equivalence}

We establish e-graphs as a persistent compiler abstraction and enable efficient exploration of equivalent programs in three steps:
introducing first-class e-graph \ac{ir} operations (\autoref{subsec:equivalence_dialect}), implementing \ac{es} using established compiler primitives (\autoref{subsec:equivalence}),
and lifting existing compiler analyses to work on e-classes (\autoref{subsec:eclass_analyses}).
Together, these steps form a complete \ac{es} framework
natively embedded in the compiler, allowing us to retain equivalences across
transformations and abstraction levels.

\subsection{Encoding E-Graphs as \ac{ir} Operations}
\label{subsec:equivalence_dialect}

\begin{listing}
  \begin{minipage}[t]{0.65\linewidth}
  \semismall
  \begin{xdsl}
|\gray{func.func @f(}\textcolor{pairedTwoDarkBlue}{\%a}\gray{) \{}|
|      \textcolor{black}{\%egraph = equivalence.graph \{}|
|  \   \textcolor{pairedFourDarkGreen}{\%two = arith.constant 2 }|
|  \   \textcolor{pairedSixDarkRed}{\%mul = arith.muli }\textcolor{pairedTwoDarkBlue}{\%a}\gray{, }\textcolor{pairedFourDarkGreen}{\%two}|
|  \    \textcolor{black}{equivalence.yield} \textcolor{pairedSixDarkRed}{\%mul}|
|  \textcolor{black}{\}}|
|  \gray{func.return \textcolor{black}{\%egraph}}|
|\gray{\}}|
  \end{xdsl}
\end{minipage}
\begin{minipage}[t]{0.27\linewidth}
  \vspace{4mm}
  \centering
  \includegraphics[width=1.27cm]{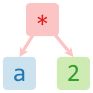}
\end{minipage}
\centering
\semismall{}
\vspace{-10pt}
\begin{center}
\begin{tikzpicture}
\draw[-{Triangle[length=5pt,width=8pt]}, pairedNegOneLightGray, line width=4pt] (0,0.4) -- (0,0);
\end{tikzpicture}
\end{center}
\vspace{-13pt}
\begin{minipage}[t]{0.65\linewidth}
  \vspace{1mm}
  \begin{xdsl}
|\gray{\%egraph = equivalence.graph \{}|
|  \textcolor{pairedFourDarkGreen}{\%two = arith.constant 2 }|
|  \textcolor{pairedSixDarkRed}{\%mul = arith.muli }\textcolor{pairedTwoDarkBlue}{\%a}\gray{, }\textcolor{pairedFourDarkGreen}{\%two}|
|  \textcolor{black}{\%one = arith.constant 1}|
|  \textcolor{black}{\%shl = arith.shli }\textcolor{pairedTwoDarkBlue}{\%a}\gray{, }\textcolor{black}{\%one}|
|  \textcolor{classBrown}{\%c = equivalence.class }\textcolor{pairedSixDarkRed}{\%mul}\gray{, }\textcolor{black}{\%shl}|
|  \gray{equivalence.yield }\textcolor{classBrown}{\%c}|
|\gray{\}}|
  \end{xdsl}
\end{minipage}
\begin{minipage}[t]{0.27\linewidth}
  \vspace{6mm}
  \centering
  \includegraphics[width=1.8cm]{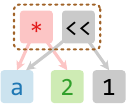}
\end{minipage}

  \caption{E-graphs are encoded as first-class \ac{ir} constructs.}
  \label{list:equivalence_conversion}
\end{listing}

To make \ac{es} available throughout the compilation process, we introduce \ac{ir} primitives for modeling the data structures at the core of \ac{es}.
These primitives are a set of \ac{ssa} operations implemented in MLIR. Together, they form our new \mlirdialect{equivalence} dialect.

\paragraph{Example} To illustrate the new primitives, consider the \ac{ir} of the function ${\tt a \times 2}$, which multiplies its only argument by two.
This function can be represented by an expression tree consisting of a multiplication node with two children, \code{a} and \code{2}.
Its \ac{ir} can be turned into a trivial e-graph (where each class has just one element) by wrapping the function body in an \mlirop{equivalence}{graph} operation (\autoref{list:equivalence_conversion}, top). Trivial e-classes are no different from the values in the original \ac{ir}.

In this \ac{ir}, applying the rewrite ${\tt y \times 2}\rightarrow {\tt y \ll 1}$ introduces a non-trivial e-class (\autoref{list:equivalence_conversion}, bottom).
Two operations, \code{shli} and \code{constant}, are added to represent the right-hand side of the rewrite.
Instead of replacing the result of the matched left-hand side with the new result, both results are instead passed as operands to a newly inserted \mlirop{equivalence}{class} operation.
This \mlirop{equivalence}{class} operation's result replaces existing uses of the \code{muli} operation (\code{\%mul}).
In the resulting \ac{ir}, non-trivial e-classes are represented with \mlirop{equivalence}{class} operations, whose use-def edges correspond to e-graph edges.

\paragraph{The \mlirdialect{equivalence} Dialect} Our new dialect features three operations that, in combination with the use-def information encoded in an \ac{ssa}-based \ac{ir}, suffice to represent an e-graph:
\begin{description}
  \item[\mlirop{equivalence}{class}] takes one or more values (analogous to e-nodes), and produces a single result.
  \item[\mlirop{equivalence}{graph}] encompasses a graph region of code for \ac{es}. All \mlirop{equivalence}{class} operations must appear within it. Operations in this region can access values defined outside, but not vice versa.
  \item [\mlirop{equivalence}{yield}] is a terminator that closes off an e-graph. It takes as operands the results that can be accessed by operations outside of the \mlirop{equivalence}{graph}. These will correspond to the root e-classes.
\end{description}

The \mlirop{equivalence}{graph} wrapper is necessary to represent cyclic e-graphs that arise due to rewrites such as ${\tt a + 0} \rightarrow {\tt a}$ (\autoref{list:cycle}).
In typical \ac{ssa}-based \ac{ir}s, value uses can only occur after their definition.
MLIR, however, supports the concept of graph regions (\autoref{subsec:background:ssa}), where this restriction is lifted.
An \mlirop{equivalence}{graph} operation contains such a graph region, allowing cycles to occur in the use-def chains.

\begin{listing}
  \centering
\begin{minipage}[t]{0.65\linewidth}
    \vspace{1mm}
    \begin{xdsl}
|\textcolor{black}{\%egraph = equivalence.graph \{}|
|   \textcolor{black}{\%zero = arith.constant 0}|
|   \textcolor{pairedFourDarkGreen}{\%c\_a} \textcolor{black}= equivalence.class \%a, \textcolor{pairedTwoDarkBlue}{\%sum}|
|   \textcolor{pairedTwoDarkBlue}{\%sum} \textcolor{black}= arith.addi \textcolor{pairedFourDarkGreen}{\%c\_a}, \%zero|
|   \textcolor{black}{equivalence.yield }\textcolor{pairedFourDarkGreen}{\%c\_a}|
|\textcolor{black}{\}}|
    \end{xdsl}
\end{minipage}
\begin{minipage}[t]{0.27\linewidth}
    \vspace{4mm}
    \centering
    \includegraphics[width=1.45cm]{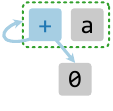}
\end{minipage}

  \caption{E-Graph cycles are modeled as graph regions.}
  \label{list:cycle}
\end{listing}

A new compiler pass embeds the e-graph in the \ac{ir} by inserting an \mlirop{equiva\-lence}{graph} operation around the operations in a \mlirop{func}{func} operation (\circled[fig:eqsat overview]{1}).
Values within the graph are left unchanged.
When rewriting adds equivalence terms, \mlirop{equivalence}{class} operations are inserted and their uses updated to reference the new class results instead of the original values.

\paragraph{Scope of our Work} To bound the scope, our implementation only supports pure, straight-line code fragments.
This proves sufficiently expressive to replicate (\autoref{sec:evaluation_herbie}) and go beyond (\autoref{sec:evaluation_rover}) existing \ac{es} applications.
Specifically, these restrictions allow the rewriting process to reorder operations freely, and replace values without worrying about side-effects or control flow.
In principle, \mlirdialect{equivalence} can also represent complex structured control flow.
For example, with the \mlirdialect{scf} dialect, which is widely used to model control flow in \ac{ir}, loops and conditionals are represented using operations with regions.
Values inside and outside those regions can be consumed by \mlirop{equivalence}{class} operations as described above.
However, applying \ac{es} on such code is non-trivial, requiring the rewriting process and extraction to be aware of scopes.
In this work, we do not explore \ac{es} on structured control flow.

\subsection{IR-Native Equality Saturation}
\label{subsec:equivalence}

\begin{figure}
  \centering
  \includegraphics[width=0.92\columnwidth]{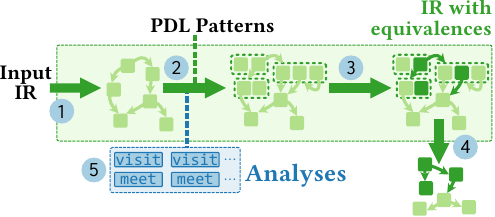}
  \caption{The steps in \ac{es} correspond to compiler passes on \ac{ir} containing \mlirdialect{equivalence} operations, using rewrite patterns specified by the user in PDL.
  Different steps include e-graph insertion (\circled{1}), e-matching and rebuilding (\circled{2}), extraction (\circled{3}, \circled{4}), and e-class analyses (\circled{5}).
  }
  \label{fig:eqsat overview}
\end{figure}

We enable efficient \ac{es} directly on an IR by (a)
reinterpreting the e-graph's invariants in terms of our \ac{ssa} encoding,
(b) extending MLIR's pattern matcher PDL to e-matching, and (c)
developing a default implementation of e-graph extraction.

\subsubsection{E-Graph Invariants in SSA}
\label{subsubsec:invariants}
\begin{figure*}
  \centering
\begin{tikzpicture}[
    font=\ttfamily\scriptsize,
    node distance=4mm and 6mm,
    stage/.style={
      draw=none,
      rounded corners=2pt,
      inner sep=4pt,
      align=left,
      fill=black!6,
    },
    label/.style={
      font=\scriptsize\bfseries,
      anchor=south west,
      inner sep=0pt,
    },
  ]

\def\circAx{-27mm}   \def\circAy{11mm}   
\def\circBx{-7mm}   \def\circBy{1mm}   
\def\circCx{-7mm}   \def\circCy{1mm}   
\def\circDx{-7mm}   \def\circDy{5mm}   

\node[stage] (s1) {
\textcolor{pairedFourDarkGreen}{\%a = class(\%x, {\normalfont\ldots})}\\
\textcolor{pairedFourDarkGreen}{\%b = f(\%a)}\\
\textcolor{pairedTwoDarkBlue}{\%c = class(\%y, {\normalfont\ldots})}\\
\textcolor{pairedTwoDarkBlue}{\%d = f(\%c)}
  };

\node[stage, right=of s1] (s2) {
\%a = class(\%x, {\normalfont\ldots})\\
\%b = f(\%a)\\
\%c = class(\textcolor{pairedFourDarkGreen}{\%a}, \%y, {\normalfont\ldots})\\
\%d = f(\%c)
  };
\node[label] at (s2.north west) {Union(\tt{\%a}, \tt{\%c})};

\node[stage, right=of s2] (s3) {
\%a = class(\%x, \textcolor{pairedFourDarkGreen}{\%y}, {\normalfont\ldots})\\
\%b = f(\%a)\\
\textcolor{black!30}{\sout{\%c = class(\%a, \textcolor{pairedThreeLightGreen}{\%y}, {\normalfont\ldots})}}\\
\%d = f(\textcolor{pairedFourDarkGreen}{\%a})
  };
\node[label] at (s3.north west) {Canonicalize};

\node[stage, right=of s3] (s4) {
\%a = class(\%x, \%y, {\normalfont\ldots})\\
\%b = f(\%a)\\
\\
\textcolor{black!30}{\sout{\%d = f(\%a)}}
  };
\node[label] at (s4.north west) {Rebuild};

\node at ($(s1.east)!0.5!(s2.west)$) {\textcolor{black!35}{\scriptsize$\blacktriangleright$}};
\node at ($(s2.east)!0.5!(s3.west)$) {\textcolor{black!35}{\scriptsize$\blacktriangleright$}};
\node at ($(s3.east)!0.5!(s4.west)$) {\textcolor{black!35}{\scriptsize$\blacktriangleright$}};

\node at ([xshift=\circAx, yshift=\circAy] s1.south east) {\circled{1}};
\node at ([xshift=\circBx, yshift=\circBy] s2.south east) {\circled{2}};
\node at ([xshift=\circCx, yshift=\circCy] s3.south east) {\circled{3}};
\node at ([xshift=\circDx, yshift=\circDy] s4.south east) {\circled{4}};

\end{tikzpicture}
\caption{We encode the relation between e-class and e-nodes, as well as the union-find data structure, directly in an SSA IR.}
\label{fig:egraph-invariants}
\end{figure*}

Central to the \ac{es} procedure are two invariants that must be maintained throughout the rewriting process~\cite{Willsey2021Egg,Nelson1980TechniquesVerification}.
First is the \emph{congruence invariant}, which ensures, roughly stated, that if $a \equiv b$, then $f(a) \equiv f(b)$.
Failing to maintain this invariant means that some implied equivalences are not represented in the e-graph, and thus cannot be exploited by rewrites.
Second is the \emph{hashcons invariant}, which says that, for every canonical e-node, there is a direct mapping to the class that contains it.
Our embedding of e-graphs in \ac{ssa} uses the use-def relationship for this instead of a hashcons, and only uses the hashcons for checking whether a node is already present.

\paragraph{Efficient Value-to-Class Mapping}
Tamagoyaki's \ac{ssa} \ac{ir}-based e-graph representation already encodes some of the relations that need to be maintained in the use-def information of the \ac{ir}.
First, the relation mapping an e-class to the e-nodes it contains is trivially represented by the operands of the \mlirop{equivalence}{class} operation representing that e-class (\circled[fig:egraph-invariants]{1}).
Second, the mapping from e-node to e-class is stored in the use-list of the value representing the e-node.
In MLIR, this use-list is stored with each value as a linked list of pointers to the operations that use it as an operand.
A key insight here is that, if an SSA value is consumed by an \mlirop{equivalence}{class} operation, then no other operation is allowed to consume it directly.
This enables $\mathcal{O}(1)$ lookups of the e-class of a given e-node by visiting the value's user, without querying a separate hashcons data structure.
Lastly, merging e-classes requires updating e-nodes that referenced the merged class.
Other frameworks capture these with a \emph{parent list} for each e-class.
In Tamagoyaki, this role is filled by the use-list of the SSA value representing the e-class.

\paragraph{A Union-Find in \ac{ssa}}
As opposed to traditional e-graph representations that use a union-find data structure to store an equivalence relation over its e-classes, Tamagoyaki stores the relation directly in the IR.
In typical e-graphs the union-find stores, for each class, a pointer to its \emph{representative class}.
When two classes are merged, this pointer is simply updated to point to the same representative.
Equivalence of two classes is verified by following their pointer chains to a common representative.
In Tamagoyaki, we allow the first operand of a class to be the result of another \mlirop{equivalence}{class} operation, enabling a direct representation of the union-find structure in the \ac{ir} (\circled[fig:egraph-invariants]{2}).

\paragraph{Rebuilding}
When existing e-classes are merged, typically because of a rewrite application, the congruence invariant is broken and needs to be restored. For example, \textbf{Union}(\code{\%a}, \code{\%c}) makes \code{f(\%a)} and \code{f(\%c)} become equivalent (\circled[fig:egraph-invariants]{2}).
Merging classes and restoring congruence after each individual rewrite application quickly becomes expensive.
To avoid duplicating work, \citeauthor{Willsey2021Egg}~\cite{Willsey2021Egg} introduced \emph{rebuilding} in which classes to be merged are added to a worklist which gets deduplicated and processed once at the end of each rewriting iteration.

Like egg, Tamagoyaki first canonicalizes e-nodes.
The class-to-value mapping (parent list) is restored, replacing the children e-classes of each e-node by their representative e-classes.
For example, if \code{\%a} and \code{\%c} are merged with \code{\%a} as representative, all users of \code{\%c} are updated to reference \code{\%a} instead (\circled[fig:egraph-invariants]{3}).
This canonicalization also restores the e-node to e-class mapping, moving all operands of \code{\%c} to its representative e-class, \code{\%a}, and erasing e-class \code{\%c}.
The next step is analogous to the congruence restoration procedure used by egg (\circled[fig:egraph-invariants]{4}).
Together these steps restore the hashcons and congruence invariants in the \ac{ir}.

\paragraph{Performance Implications}

Maintaining the e-graph invariants directly in \ac{ssa} \ac{ir} does come at a price.
Compared to traditional e-graph implementations, Tamagoyaki typically performs more work during class creation and rebuilding.
The former is because we do not insert a new class immediately when a new operation is inserted in the IR, but only after some pattern merges values that lack a class.
This means the original operation can already have users that need to be updated to reference the new class.
Similarly, during rebuilding, if two operations have become identical due to a class merge, we must explicitly remove one and update all its users to reference the other, while in a traditional e-graph, the hashcons simply prevents the second operation from being inserted in the first place.
While this additional cost is a disadvantage, for the practical use cases we evaluate (Sections \ref{sec:evaluation_herbie} \& \ref{sec:evaluation_rover}) we found the benefit of persisting equivalences in \ac{ir} outweighs the extra overhead.

\subsubsection{Compiler-Native E-Matching via PDL}
\label{subsubsec:ematching}

\begin{figure}
  \centering
  \includegraphics[width=\linewidth]{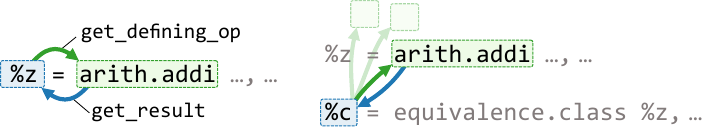}
  \caption{(left) In MLIR's default, destructive pattern rewriting framework, the \mlirop{pdl\_interp}{get\_result} and \mlirop{pdl\_interp}{get\_defining\_op} operations are each other's inverse. (right) In \mlirdialect{equivalence}, this is not the case because each value comes from one of multiple equivalent operations.}
  \label{fig:pdl_interp}
\end{figure}

We extend MLIR's Pattern Description Language (PDL) (\autoref{subsec:background:pdl}) infrastructure to enable compiler-native e-matching (\autoref{subsec:background:equivalence}).
Our implementation allows users to run existing PDL patterns on code extended with equivalences.
We adapt the \mlirdialect{pdl} to \mlirdialect{pdl\_interp} lowering (\autoref{list:pdl}) using xDSL~\cite{xdsl}, extending the output with new e-match-specific operations.
These additions enable e-matching without modifying existing destructive rewriting infrastructure.

PDL is not designed to match patterns in an e-graph, as the indirections caused by our newly introduced \mlirdialect{equivalence} operations are not taken into account (\autoref{fig:pdl_interp}).
We resolve this by introducing \mlirdialect{ematch}, a dialect that contains operations that model e-matching and maintain e-graph invariants.
Consider the example rewrite: \code{a \& a} $\rightarrow$ \code{a}~(\autoref{list:pdl}).
The standard lowering creates a \mlirop{pdl\_interp}{replace} operation to \emph{replace} \code{(a \& a)} by \code{a}, losing the existing expression.
In contrast, \ac{es} merges the e-class containing \code{(a \& a)} and the e-class containing \code{a}.
We implement this with an \mlirop{ematch}{union} operation, merging the two e-classes non-destructively.

Although MLIR's use-list avoids the need to query a hashcons to get an e-node's containing e-class, Tamagoyaki still maintains a separate hashcons to ensure that no operation appears in the e-graph twice.
In the new lowering every \mlirop{pdl\_interp}{create\_operation} is followed by an \mlirop{ematch}{dedup}.
Each \mlirop{ematch}{dedup} queries the hashcons to check if the e-graph contains an identical operation, and reuses it if so.
We note that the ability to de-duplicate during rewriting may also be useful for regular, destructive rewriting.
For the hashcons we reuse data structures from existing passes, e.g., \ac{cse}.
These and further modifications not covered here add first-class support for non-destructive rewriting to MLIR.

To make MLIR native e-matching efficient, the new lowering adapts the order in which constraints are checked to avoid iterating over an e-class multiple times.
The unique challenge in e-matching is that every time we query a value's defining operation (normally a cheap query) we need to iterate over all operations within an e-class.
Therefore, our lowering groups constraints together to exit a matching routine as early as possible in order to avoid redundant iterations.

One advantage of PDL is that it can match multiple patterns in a single routine. To let it perform efficient multi-pattern e-matching, again some modifications are required.
As noted by \citeauthor{DeMoura2007EfficientSolvers}~\cite{DeMoura2007EfficientSolvers}, e-matching code for independent patterns should be separated.
Their solution achieves this using a special \code{choose} operation.
Our adapted \mlirdialect{pdl} achieves the same by generating nested \mlirop{pdl\_interp}{for\_each} operations that loop over the e-classes' operands to implement iteration over equivalences.
When a pattern fails to match for all equivalences, the matcher breaks out of the loop, explicitly discarding the equivalences that were explored, and jumps to the next pattern.
This modification is key to enabling efficient multi-pattern e-matching.

Without the two described adaptations, the generated matching code fails to terminate within an acceptable time even for a small pattern set.
Combined, the two adaptations enable efficient e-matching with unmodified \mlirdialect{pdl} patterns.

\subsubsection{Extraction from an \mlirop{equivalence}{egraph}}
\label{subsubsec:extraction}

Since the e-graph is embedded in the IR, extraction is simply a case of replacing each \mlirop{equivalence}{class} by one of its operands, updating its uses and erasing any unused operands.
We separate extraction into two passes.
First, the \emph{selection} pass adds an attribute to each \mlirop{equivalence}{class} indicating which of its operands yields the lowest cost according to a given cost model (\circled[fig:eqsat overview]{3}). The \emph{replacement} pass then substitutes each \mlirop{equivalence}{class} with the selected operand. The unselected e-class operands are removed from the program (\circled[fig:eqsat overview]{4}), leaving only the optimal program in the IR.

If the e-graph contains cycles, the extracted program's order may violate \ac{ssa} use-def order.
If this is not acceptable, a topological sort of the operations can be carried out to restore \ac{ssa} order.
As we assume the program consists of pure operations (\autoref{subsec:equivalence_dialect}), reordering them is safe.

Tamagoyaki users may either develop their own custom extraction passes or use Tamagoyaki's defaults that greedily extract the lowest cost operand for each \mlirop{equivalence}{class}.
Using the default approach, costs for each type of operation can be specified using a configuration file or through the use of attributes on individual operations (\autoref{sec:evaluation_herbie}).
However, it is also simple to prune an \mlirop{equivalence}{egraph} as it is just IR.
A custom replacement pass may prune operands from all \mlirop{equivalence}{class} operations or just replace a subset of them. As a result, Tamagoyaki lets users extract an optimal program according to a multitude of cost models, e.g., using one cost model to prune then another to break ties (\autoref{sec:evaluation_rover}).

\subsection{Seamless Integration of Analyses}
\label{subsec:eclass_analyses}

Tamagoyaki allows for the seamless lifting of existing MLIR program analyses to perform e-class analysis. Specifically, e-class analysis in Tamagoyaki can readily reuse existing MLIR dataflow analyses and their lattices. Moreover, those dataflow analyses readily operate on IR containing e-classes.

\paragraph{Interfaces to be Coupled}
E-class analyses track and propagate abstract elements from a lattice through the e-graph~\cite{Willsey2021Egg}.
For example, an interval analysis can determine the interval in which the e-nodes in an e-class fall when evaluated~\cite{Coward2023CombiningInterpretation}.
Egg introduced the e-class analysis interface~\cite{Willsey2021Egg}, consisting of three functions: (1)
 \code{make} initializes analysis data for a newly created e-node, (2) \code{merge} combines analysis data (using the lattice meet or join operation) of two e-classes when they are unioned, and (3) \code{modify} modifies an e-class based on its analysis data.

By contrast, MLIR implements program analyses via a sparse dataflow analysis framework and a fixed point solver that allows developers to implement analyses that propagate information through the \ac{ir}.
Crucially, for a sparse forward analysis, where information is propagated from an operation's operands to its results, the developer is required to implement two functions (\circled[fig:eqsat overview]{5}):
(1) \code{visit} initializes the analysis data for an operation result based on its operands, and (2) \code{meet} implements the meet of the analysis lattice, used primarily when merging control flow paths.

\paragraph{Coupling the Interfaces}
We define the \code{visit} method for \mlirop{equivalence}{class} as the \code{meet} of its operands' analysis data. With this definition, the dataflow solver
treats \mlirop{equivalence}{class} operators just like any other operator. This means
that we can, without modification, run dataflow analyses on \ac{ir} containing e-classes, even after \ac{es} has terminated.
At the same time, we replicate egg's e-class analysis by implementing the \code{make} and \code{merge} functions using the \code{visit} and \code{meet} methods, respectively.
This means that existing dataflow analysis lattices can be reused in e-class analysis without additional effort.
By reusing existing dataflow analyses, we can leverage the existing analysis infrastructure in MLIR, such as the data structures for representing analysis data.
Lastly, whilst the dataflow analysis framework does not have an analogue to \code{modify}, its functionality can be replicated by a compiler pass that modifies the \ac{ir} based on the analysis data.

We believe that this seamless integration of analyses will be a key advantage of our approach, as it allows for the reuse of existing analyses and their lattices, and enables the combination of multiple analyses to improve their precision.
For now, this integration in Tamagoyaki further expands the reusability of MLIR compiler infrastructure.

\section{Case Study 1: Floating-Point Accuracy}
\label{sec:evaluation_herbie}

Through the following case study we will demonstrate how we can leverage Tamagoyaki\footnote{For evaluation, we use the MLIR implementation of Tamagoyaki.} to reproduce a subset of Herbie~\cite{Panchekha2015Herbie}, a project that uses \ac{es} to automatically improve the numerical stability of floating-point expressions.
Herbie was chosen for its sophisticated use of \ac{es}, its consistent role in evaluating new advances in \ac{es}~\cite{Willsey2021Egg,Zhang2023Egglog}, and its demonstrated value when deployed within a compiler~\cite{qian2026poseidon}.

Given an expression, Herbie uses egg (and other techniques)
to construct a set of alternative candidates that are equivalent when evaluated
using real arithmetic.
From this set Herbie selects the candidate with the lowest numerical error based
on random sampling of floating-point inputs.

By default, Herbie runs multiple rounds of refinement to arrive at a final result.
Additionally, Herbie implements a number of optimizations outside of the core procedure. \emph{Regime inference} attempts to split the input domain of the expression to create a piecewise function with lower error than a single expression. \emph{Series expansion} produces additional candidate expressions by approximating existing expressions using Taylor series.
We focus solely on the core steps of Herbie's procedure, which consists of sampling points, the \ac{es} phase, and selecting the optimal candidate by evaluating their floating-point accuracy.
To this end, we implement a compiler pass using Tamagoyaki that we compare against a single round of Herbie's main loop, with series expansion and regime inference disabled.

\begin{figure}
  \centering
  \includegraphics{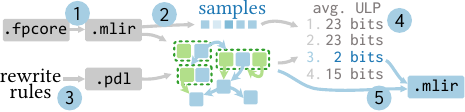}
  \caption{Tamagoyaki can implement Herbie's core optimizations as MLIR-native non-destructive rewriting.}
  \label{fig:herbie flow}
\end{figure}

In this flow, a parser maps operators of Herbie's input language, FPCore~\cite{Damouche2017TowardAnalysis}, to the \mlirdialect{arith} and \mlirdialect{math} dialects (\circled[fig:herbie flow]{1}).
The translated pure, straight-line function body is then wrapped in an \mlirop{equivalence}{graph} operation.
Additionally, 256 sample inputs are generated for the program, guided by an interval analysis to avoid generating samples that lead to errors such as division by zero (\circled[fig:herbie flow]{2}).
The analysis uses Rival3~\cite{flatt2023making}, an interval analysis library written in Rust that is also used by Herbie.

For each sample input, the \emph{ground-truth} result of the program is computed.
This is done by evaluating the program at high precision, again using the Rival3 library, which uses MPFR~\cite{mpfr2007} under the hood, in addition to many optimizations such as dynamically increasing precision.

For the \ac{es}, we converted Herbie's set of rewrite rules, which are implemented in Racket, to \mlirdialect{pdl} rewrite patterns (\circled[fig:herbie flow]{3}).\footnote{We evaluated against Herbie commit \texttt{5500c96}.}
Herbie contains rules that expand any term to an expression, e.g., \code{x} $\rightarrow$ \code{log(exp(x))}.
Such rules are not directly expressible in \mlirdialect{pdl}, so we omit them in our Herbie re-implementation, as they had no significant impact on the accuracy of the generated expressions.

After lowering the \mlirdialect{pdl} patterns to \mlirdialect{pdl\_interp}, Tamagoyaki runs \ac{es}.
Following Herbie's default settings, we set no limit on the number of \ac{es} iterations, but limit the e-graph to a total of 4000 e-nodes.

At this point in its flow, Herbie extracts \emph{patches} from the e-graph.
These are all the expressions that have been inserted in the e-classes containing the original expression's sub-expressions.
Herbie does this by first converting egg's e-graph into an e-graph data structure implemented in Racket on which it can run a custom extraction procedure.
For each of these patches, a \emph{candidate} is then constructed by patching the patch into the original expression.
These candidates are then evaluated on the sample inputs (not with the high-precision library, but using regular floating-point arithmetic) and compared against the ground-truth values to compute their errors.
That comparison is done by computing the \ac{ulp} distance, which is the number of representable floating-point values between the two values.

Unlike Herbie, Tamagoyaki does not have to convert the e-graph first or extract candidate expressions from the e-graph explicitly.
Since the e-graph is not stored externally in a framework like egg, but rather directly in the \ac{ir}, Tamagoyaki can compute the required values directly in the compiler pass (\circled[fig:herbie flow]{4}).
This is facilitated by the decoupling of extraction into selection and replacement (\autoref{subsubsec:extraction}).
Tamagoyaki tracks the operations from the original program throughout \ac{es}.
The original program serves as a baseline against which mutations are compared.
Tamagoyaki runs an initial greedy cost model to select the cheapest values in the classes that do not contain an operation from the original program, such that each class now has a selected operand.
Tamagoyaki then runs batched constant propagation: each of the greedily selected SSA values that is a dependency of the program result value, as well as all the equivalent values in the classes that contain an operation from the original program, are evaluated one by one (topologically sorted) on all the samples at once.

Finally, we are left with $\text{\#samples}\times{}\text{\#patches}$ result values, where each patch corresponds to a deviation from the original greedy selection in a single class.
These local costs are then used to make a final greedy selection and replacement (\circled[fig:herbie flow]{5}).
For a fair comparison between Herbie's generated expression and our own, we convert the final optimized code back to FPCore.
This allows us to evaluate both expressions using the same set of random samples, computing accuracy as the average proportion of bits that match the ground-truth result across all samples.

\paragraph{Accuracy Results}

\begin{figure}
  \centering
  \includegraphics{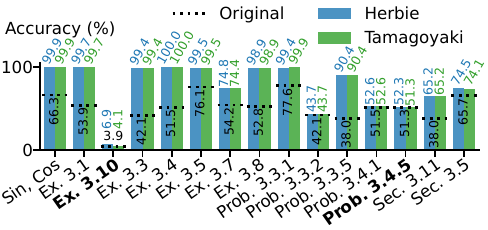}
  \caption{Herbie (and Tamagoyaki) improve accuracy on 15 out of 32 total benchmarks. Tamagoyaki is within 99\% of Herbie's accuracy except for 2 benchmarks. For those, Tamagoyaki's output is never worse than the original program.}
  \label{tab:accuracy_comparison}
\end{figure}

We evaluate our re-implementation using the 31 Herbie-contributed benchmarks in the FPBench suite\footnote{https://fpbench.org/benchmarks.html\#from\%3Aherbie} and compare the accuracy of the original input expression to the optimized Herbie output and the output optimized with our implementation (\autoref{tab:accuracy_comparison}).
Our re-implementation is able to match Herbie's accuracy on all but two benchmarks.
For all benchmarks, \ac{es} is aborted due to the e-node limit being reached.
This means that the order in which rewrites are applied \emph{can} have impact on the final optimized program.
We believe that a different ordering of rewrites would lead to other rules being applied, leading to a more accurate result by chance.
These results demonstrate that our Tamagoyaki-based implementation of the \mlirdialect{equivalence} dialect is sufficiently powerful to replicate the core mechanism of Herbie.

\paragraph{Compilation Time}
\begin{figure*}[b]
  \centering
  \includegraphics{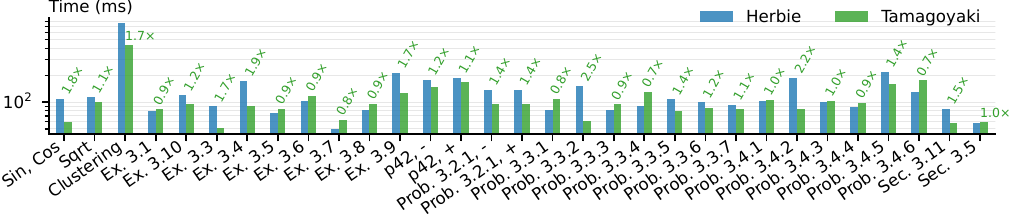}
  \caption{Our implementation with Tamagoyaki has a geomean speedup of $1.18\times$ over Herbie.}
  \label{fig:timing_comparison}
\end{figure*}
We ran our evaluation on a system with an AMD Ryzen 9 9950X CPU (16 cores/32 threads), using only a single core for all experiments.
We compiled Tamagoyaki in release mode, linking against MLIR commit {\tt d08bb68}.
For Herbie, we used Racket v8.15 and cargo 1.91.0 to build egg.
For the comparison against Herbie, we ran our implementation with all rewrite patterns combined into a single \mlirdialect{pdl\_interp} matcher.
Our timings of Tamagoyaki do not include the time spent on loading the \mlirdialect{pdl\_interp} patterns, as this step currently includes parsing and a liveness analysis to generate the executable bytecode.
In the future it should be possible to virtually eliminate this overhead by serializing the generated bytecode and loading it directly.
For Herbie itself, we report the time as reported by Herbie's built-in timing infrastructure.
Since Herbie samples points to be used during optimization (256) as well as during the final evaluation (8000), we multiply the total sampling time by $256/8256$ to approximate the time spent on sampling for optimization only, which is what we compare against.

Overall, we reach similar compilation time as Herbie on most benchmarks (\autoref{fig:timing_comparison}), with a geomean speedup of $1.18\times$.
If we include the time loading the \mlirdialect{pdl\_interp} patterns, we see a \emph{slowdown} to $0.78\times$ of Herbie's performance.
Looking at \ac{es} (matching, rewriting, and rebuilding) in isolation, we measure a slowdown of $0.78\times$.
This can be partially attributed to the fact that Tamagoyaki performs more work during class creation and rebuilding than traditional e-graph implementations, as described in \autoref{subsubsec:invariants}.
\begin{table*}
  \caption{The Multi-Level flows deliver the greatest Delay (ns) and Area ($\mu\text{m}^2$) improvements. We \textbf{bold} the best in each category. Opt. Time (ms) is the time for \ac{es} and for the additional CIRCT passes applied to the e-graph in the last flow.}
  \label{tab:rover_eval}
  \begin{tabular}{l r r r r r r r r r r r}
    \toprule
    \multirow{2}{*}{Benchmark} & \multicolumn{2}{c}{No EqSat} & \multicolumn{3}{c}{Single-Level} & \multicolumn{3}{c}{\textbf{Multi-Level}} & \multicolumn{3}{c}{\textbf{Multi-Level + CIRCT Passes}}                                                                         \\
                               & Area                 & Delay                                    &Area            & Delay & Opt. Time      & Area & Delay & Opt. Time & Area & Delay & Opt. Time     \\

    \cmidrule(lr){1-1} \cmidrule(lr){2-3} \cmidrule(lr){4-6} \cmidrule(lr){7-9} \cmidrule(lr){10-12}
FirFilter & \textbf{230} & 963 & 330 & \textbf{674}& 4 & 330 & \textbf{674} & 4 & 330 & \textbf{674} & 10 \\
Adpcm & 159 & 357 & \textbf{89} & \textbf{286}& 6 & \textbf{89} & \textbf{286} & 5 & \textbf{89} & \textbf{286} & 13 \\
ShiftedFma & 910 & 819 & 910 & 819& 4 & \textbf{857} & \textbf{727} & 4 & \textbf{857} & \textbf{727} & 17 \\
ShiftMult & 1339 & 741 & 825 & 771& 4 & 1339 & 741 & 7 & \textbf{775} & \textbf{687} & 38 \\

  \end{tabular}
\end{table*}

\paragraph{Speedup of Combining Patterns}
We also measure the performance improvements when matching with a single matcher function that includes all patterns compared to matching each pattern individually on the e-graph.
This functionality is enabled by building on MLIR's \mlirdialect{pdl} and its lowering to \mlirdialect{pdl\_interp}, which generates matching code that matches multiple patterns at once.
We only measure the time spent during actual matching, excluding other steps such as rewriting and rebuilding.
We see a geomean speedup of $2.38\times$, showing that the combined pattern e-matching leads to significant performance improvements.

\paragraph{Lowering Barriers}
Herbie is a powerful tool, allowing software developers to improve the numerical stability of procedures with floating-point arithmetic.
In order to use it in a compilation workflow, the developer must initially identify potentially
unstable mathematical expressions in their code and implement conversions to FPCore,
including any relevant input constraints, as well as conversions back to their input \ac{ir} for the optimized results.
A re-implementation of Herbie embedded in the compiler allows its techniques to be conveniently applied within compilation pipelines with minimal effort for the developer.

\section{Case Study 2: Circuit Design}
\label{sec:evaluation_rover}

Building on the motivational example (\autoref{sec:motivating}), we developed a circuit optimization flow based on CIRCT~\cite{Eldridge2021MLIRInfrastructure} that exploits the power of multi-level \ac{es} and a persistent e-graph representation.
Our multi-level circuit optimization flow reduces circuit delay across a set of benchmarks even more than a single-level \ac{es} optimizer, as explored in the ROVER project~\cite{Coward2024ROVER:Rewriting}.
More recent work~\cite{kong2026improving}, translated further hardware abstractions across to egglog, an external library~\cite{Zhang2023Egglog}, to expand the scope of \ac{es} in hardware optimization. 
In contrast, we embed \ac{es} natively in a mature hardware compilation framework.

Given a Verilog design, ROVER constructs an e-graph representation of the design using a custom language.
ROVER then uses egg to perform \ac{es} with a set of bitvector rewrites.
An optimal design is extracted from the e-graph based on circuit area~\cite{Coward2024ROVER:Rewriting} and delay~\cite{Coward2024Constraint-AwareOptimization} cost metrics.

Since ROVER is not open-source and was evaluated using proprietary circuit design flows, a direct comparison is beyond our reach.
Instead we replicate ROVER using Tamagoyaki to explore CIRCT's \mlirdialect{comb} dialect as it targets the same abstraction level as ROVER's custom representation.
A subset of ROVER's rules are implemented in \mlirdialect{pdl} and modified to operate over \mlirdialect{comb}.
We note that CIRCT lacks a dialect that precisely matches the semantics of ROVER's representation, which supports operators with operands of different widths, whilst \mlirdialect{comb} requires all operands to be the same width.
Therefore we do not claim an exact replica, instead calling this our ``Single-Level'' flow (\circledgreen[fig:circt_flows]{a}).
This flow takes a design in CIRCT \ac{ir} and leverages Tamagoyaki to perform \ac{es} using the ROVER rewrites, constructing an e-graph of equivalent \mlirdialect{comb}-level designs.

To exploit Tamagoyaki's capabilities, our ``Multi-Level'' flow explores an additional level of abstraction by implementing CIRCT's \mlirdialect{comb} to \mlirdialect{datapath} lowering pass as \mlirdialect{pdl} rewrites.
We saw this lowering for a multiplier earlier (\autoref{sec:motivating}).
Our ``Multi-Level'' flow again takes a design in CIRCT IR and uses Tamagoyaki to perform \ac{es} with \emph{both} the lowering rewrites and the ROVER rewrites, before moving straight to extraction (\circledgreen[fig:circt_flows]{a}+\circledgreen{b}).

A final flow, ``Multi-Level + CIRCT Passes'', applies two existing CIRCT passes directly to the e-graph generated by ``Multi-Level'', rather than moving straight to extraction (\circledgreen[fig:circt_flows]{a}+\circledgreen{b}+\circledgreen{c}).
First, \code{canonicalize} applies simplifications (e.g., $a+a \rightarrow a \ll 1$) defined for the \mlirdialect{comb} and \mlirdialect{datapath} dialects. Second, \code{comb-int-range-narrowing} performs width reduction, e.g.,
$
  \code{zext}(a) \mathbin{\&} \code{zext}(b) \rightarrow \code{zext}(a \mathbin{\&} b).
$
Exploiting the persistent representation, the passes destructively replace operators in the e-graph, which simplifies the extraction process as each CIRCT operator is canonicalized.

The three Tamagoyaki-based flows use the same extraction procedure, guided by cost models for circuit area and delay.
All flows perform an initial \emph{pruning} extraction, retaining only the designs that achieve the lowest delay.
Tie-breaks are resolved via a second extraction pass based on the circuit area cost model, producing a single circuit design.
The generated designs are synthesized using \code{circt-synth}, then mapped to the ASAP 7nm technology library~\cite{Clark2016ASAP7:Kit} using ABC~\cite{Brayton2010ABC:Tool} to derive area and delay metrics.

\begin{figure}
  \centering
  \includegraphics{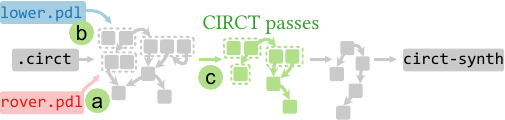}
  \caption{Tamagoyaki-based circuit optimization flows.}
  \label{fig:circt_flows}
\end{figure}

To evaluate, we take four designs from ROVER's benchmark suite~\cite{Coward2024ROVER:Rewriting}.
The other ROVER benchmarks are either closed-source designs or constant-multiplication designs that are better handled by custom tools~\cite{Kumm2016MultipleArrays}.
We compare four circuit optimization flows across these four benchmarks (\autoref{tab:rover_eval}).
The baseline ``No EqSat'' flow passes the given CIRCT IR straight to \code{circt-synth}.
The ``Single-Level'', ``Multi-Level'' and ```Multi-Level + CIRCT Passes''' flows each perform four iterations of rewriting.

All Tamagoyaki-based flows improve both circuit delay and area when compared against the traditional ``No EqSat'' synthesis tool.
The ``Multi-Level'' flow matches the ``Single-Level'' flow on two benchmarks (FirFilter \& Adpcm) and reduces area by 6\% and delay by 11\% on ShiftedFma.
The improvement can be attributed to the multi-level exploration that reveals new designs not representable using \mlirdialect{comb}.
The ``Multi-Level + CIRCT Passes'' flow matches the ``Multi-Level'' flow on all but ShiftMult, where the CIRCT passes help guide extraction to the best design overall.
To evaluate the computational cost of accessing these gains we compare the run times of the four iterations of e-graph rewriting.
On average, ``Multi-Level'' is 15\% slower than ``Single-Level'', which is expected as we explore a larger e-graph.
Interestingly, the CIRCT passes add significantly to the optimization time, likely due to the fixpoint loop performed in the extra passes.

We conclude that Tamagoyaki's e-graph persistence over multiple abstraction levels can indeed yield additional optimizations. 
Furthermore, both case studies combined confirm that the infrastructure is reusable across very different domains and optimization objectives.

\section{Related Work}
\label{sec:related}

The connection between compilers and \ac{es} has been explored in
numerous works, each picking a point on the spectrum from fully custom to fully
general integration. In fact, the first work to propose \ac{es}
demonstrated its capabilities via a Java bytecode
optimizer~\cite{Tate2009EqualityOptimization}.
Since then the leading standalone \ac{es} libraries,
egg~\cite{Willsey2021Egg} and egglog~\cite{Zhang2023Egglog}, have been used to
develop numerous compiler and code optimization
applications~\cite{Panchekha2015Herbie, Coward2024ROVER:Rewriting, Cheng2024SEER:MLIR,Merckx2026EqualitySaturationOptimizing}.
Both libraries introduced new techniques to improve the performance of equality
saturation. First, egg introduced rebuilding and e-class analysis (\autoref{subsec:background:equivalence}).
Then egglog, inspired by datalog, reimagined the underlying data structures
used to represent the e-graph to greatly improve performance.
Each application built on these general purpose libraries, develops translations
to/from source code to custom \acp{ir} that can be transformed by the chosen
library.
As shown in our evaluation (\autoref{sec:evaluation_herbie}), applications built on Tamagoyaki offer comparable performance and can take advantage of existing compiler infrastructure.

Two works sought to improve MLIR integration~\cite{dialegg2025,vohra2025mind}.
The first developed a bespoke \ac{es} pass using egg to optimize machine learning code~\cite{vohra2025mind}.
The second, more widely applicable approach, developed a framework for representing any MLIR dialect in egglog~\cite{dialegg2025}.
In this framework users must still define their own \ac{es} rewrites and cost functions to replace a single compiler pass.
Their approach represents a middle ground, lowering the development overhead and benefiting from a performant \ac{es} implementation.
The downside is that each time the compiler calls out to egglog and returns, the entire e-graph is discarded meaning that the learned equalities cannot be accessed by subsequent passes.

The deepest embedding of \ac{es} in a compiler was explored in the Cranelift
compiler~\cite{cranelift}. Unlike the works discussed so far,
the Cranelift developers do not utilize a specialized \ac{es} library.
Instead they develop the \ac{aegraph} that directly represents Cranelift's \ac{ir}
in conjunction with a \ac{cfg} skeleton that stores each function's \ac{cfg}.
The skeleton lets Cranelift represent control flow, but does not permit control flow rewrites, limiting the scope of optimizations.
The \ac{aegraph} developers selectively discarded certain features of \ac{es}
to limit the compile time and memory overhead associated with \ac{es}.
Whilst similar to \mlirdialect{equivalence}, we retain more of the generality of
libraries like egg and by embedding the e-graph in MLIR  we open up the possibility
of rewriting across different abstraction levels. Also, since MLIR uses structured
control flow constructs, every function consists of a single basic block,
and control flow is captured by operations containing nested code regions.
Conceptually, this can allow control flow to be rewritten just the same as other
operations. Lastly, by leveraging graph regions we can represent cyclic
e-graphs which Cranelift's \acp{aegraph} cannot, allowing Tamagoyaki to execute full
\ac{es}.

\section{Conclusion}

By embedding the e-graph directly in the compiler's \ac{ir}, we eliminate any
translation to and from external libraries and can build an e-graph that persists throughout compilation phases.
The persistent e-graph facilitates \ac{es} across multiple abstraction levels helping to mitigate the phase-ordering problem.
We have shown that existing compiler infrastructure such as MLIR's \mlirdialect{pdl} dialect, graph regions, and dataflow framework can be harnessed to more easily implement \ac{es}.
Even though \ac{es} in Tamagoyaki currently only works for straight-line, pure code, we show that this already suffices in different domains.
We used Tamagoyaki to replicate the functionality of a floating-point rewriting tool, demonstrating the flexibility of the approach.
Furthermore, we showed that by combining Tamagoyaki with CIRCT, we are able to discover optimized circuit designs that are beyond the reach of existing, single-level circuit optimizers that leverage \ac{es}.
We believe the concepts introduced in this paper can serve as a foundation for our future work that will expand the scope of equivalence-based optimizations to more complex code with control flow and side effects.
We expect that Tamagoyaki will unlock a variety of new applications, offering an easy way for the compiler community to explore the advantages of \ac{es}.


\bibliography{references}

\begin{draftonly}
  \cleardoublepage
  \appendix
  \section{Formatting and Writing Guidelines}

  These formatting guidelines aim to standardize our writing. They ensure that
  papers with multiple authors have a consistent look and that commonly occurring
  items are formatted in ways that are known to work well.

  \subsection{Figures}
  \label{appendix:figures}

  \paragraph{Referencing Figures} When referencing figures from the text we
  ensure the following:
  \begin{description}
    \item [All figures are referenced] A paper with un-referenced figures appears incomplete.
          We can check this using the \texttt{refcheck} package.
          Issuing \texttt{make refcheck} on the commandline lists all \emph{labeled} elements that are not referenced in the text.
    \item [References to figures are brief and easy to skip]~\\
          We minimize the number of words needed to refer to a figure. Reducing
          the number of non-information-carrying words directly increases
          the density of interesting content. When skipping references
          becomes easy, reading quickly while ignoring figures remains a
          smooth experience. The best and briefest reference to a figure
          is a link in parenthesis that is added after the subject
          representing the content depicted in a figure:\\
          {\color{pairedTwoDarkBlue}\textit{Figure
            X shows the design of A, which consists of ...}}\\
          $\to$ {\color{pairedFourDarkGreen}
              \textit{The design of A (Figure X) consists of ...}}
    \item [The text is always self-contained without figures] ~\\ The reader
          should be able to read the text without ever looking at any
          figure. They should still understand the text and get the key
          message of each figure directly from the text. By not forcing
          the reader to analyze a figure while reading, we increase
          readability as the reader can continue reading without having
          to skip between text and figures. Such writing style also helps
          to guide the thoughts of the reader, who can (for a moment)
          trust our summary of the figure and does not need to develop
          their own interpretation on-the-fly, a task which often yields
          results that do not fit the flow of our exposition. Readers
          typically only feel that their reading is interrupted if there
          is no explanation of a figure at all. Hence, we do not need to
          discuss all details of a figure, but half a sentence that explains the
          core idea is typically sufficient for a reader to continue
          reading.  By making
          our text self-contained even when ignoring figures the reader
          experiences a smooth and uninterrupted reading experience.\\
          {\color{pairedTwoDarkBlue}
          \textit{The speedups are presented in Figure X. < a new topic> }}\\
          $\to$ {\color{pairedFourDarkGreen}\textit{Our approach outperforms the state of the art
                XXX-library (Figure 3) demonstrating more than 4x speedup on
                test case 1 and 2 and a geometric mean speedup of 1.5x over all
                20 test cases.}}
  \end{description}
  We reference figures in text using
  \texttt{\symbol{92}autoref\{fig:speedup\}} for a figure with label
  \texttt{fig:speedup}.  The use of autoref ensures that all references
  to figures are formatted consistently, e.g. as \autoref{fig:speedup}.
  \paragraph{Color Scheme}

  In Figures we use a color scheme that is print-friendly and also visible
  with red-green blindness. The following colors are all print-friendly
  and red-green save when only using Color 1-4:

  \medskip
  {
    \small
    \newcolumntype{a}{>{\columncolor{pairedOneLightBlue}}c}
    \newcolumntype{b}{>{\columncolor{pairedTwoDarkBlue}}c}
    \newcolumntype{d}{>{\columncolor{pairedThreeLightGreen}}c}
    \newcolumntype{e}{>{\columncolor{pairedFourDarkGreen}}c}
    \newcolumntype{f}{>{\columncolor{pairedFiveLightRed}}c}
    \newcolumntype{g}{>{\columncolor{pairedSixDarkRed}}c}

    \begin{tabular}{a b d e f g}
      Color 1  & Color 2  & Color 3  & Color 4  & Color 5  & Color 6  \\
      \#a6cee3 & \#1f78b4 & \#b2df8a & \#33a02c & \#fb9a99 & \#e31a1c
    \end{tabular}
  }

  We de-emphasize components in figures by using additionally two shades of gray.
  Especially in complex figures, it is often helpful to de-emphasize visual
  elements that we want to represent but that should not be the focus of a
  reader's attention.

  \medskip
  {
    \small
    \newcolumntype{h}{>{\columncolor{pairedNegOneLightGray}}c}
    \newcolumntype{i}{>{\columncolor{pairedNegTwoDarkGray}}c}

    \begin{tabular}{h i}
      Color -1 & Color -2 \\
      \#cacaca & \#827b7b \\
    \end{tabular}
  }

  Single-color graphs are plotted in Color 1 - Light Blue.

  \paragraph{Labels in Figures}
  Complex diagrams often benefit from labels inside the diagrams. We suggest to
  use a filled circle (e.g, in light blue) to highlight these numbers and use
  these references, e.g., \circled{1} implemented as \texttt{\textbackslash{}circled\{1\}}, in the text to refer to them.

  \paragraph{Captions and Core Message}
  \label{appendix:captions}

  Each figure should have a caption that makes a clear statement about this
  figure, as such a statement makes it easier for the reader to (in)validate the
  figure as evidence for the claim we make. Traditionally, figures often have a
  caption indicating its content:\\ {\color{pairedTwoDarkBlue} \textit{$\cdot$
    Speedup of approach A vs approach B on system X}}\\ {\color{pairedTwoDarkBlue}
  \textit{$\cdot$ Architecture diagram of our solution}}\\ While these statements
  clearly state the content of a figure at the meta-level, they often lack
  information about the precise content and the claim a figure is meant to
  evidence. While knowing that a figure is an architecture diagram is useful for
  the reader when looking at the figure the reader automatically asks two
  questions: (a) what properties set this architecture apart and (b) does its
  implementation deliver the claimed properties? Or, in more general terms, what
  claims do we aim to evidence with this figure and does the figure provide the
  needed evidence to support our claims? In theory, this information could be
  contained in the text of the paper, but to optimize for readers who skim the
  figures first, we want to offer them as part of the caption. Nevertheless, it
  makes often sense to word the caption strategically to still document the
  meta-level content of a figure.\\ $\to$ {\color{pairedFourDarkGreen}
      \textit{$\cdot$ Approach A is consistently faster than approach B, except for
        inputs that are not used in practice}}\\ $\to$ {\color{pairedFourDarkGreen}
      \textit{$\cdot$ The architecture of our design increases reusability by making
        components A, B, \& C independent of the core.}} For example, after reading the
  last caption the reader can validate if the architecture design indeed enables
  the promised independence, and we can double-check while drafting the paper that
  our figure is visualized to facilitate checking if it works as evidence. ! This
  does not mean we should mislead with our figure but rather make things easy to
  check. If our figure or data would not support our claim, it should be similarly
  easy to invalidate our claim!

  \subsubsection{Plots} We use matplotlib to create performance
  plots such as \autoref{fig:speedup}. We use the following
  formatting guidelines:
  \begin{itemize}
    \item Use a vertical y-label to make it easier to read.
    \item Remove top and right frames to reduce visual noise
          and allow the reader to focus on the data in the
          figure.
    \item Provide the concrete data at the top of each bar.
  \end{itemize}

  \noindent
  We also suggest to follow these technical remarks:
  \begin{itemize}
    \item Create pdf plots and do not use bitmap formats (e.g., png) to
          ensure high quality when zooming in.
    \item Avoid Type-3 bitmap fonts by
          setting fonttype to 42.
  \end{itemize}

  \begin{figure}
    \includegraphics[width=\columnwidth]{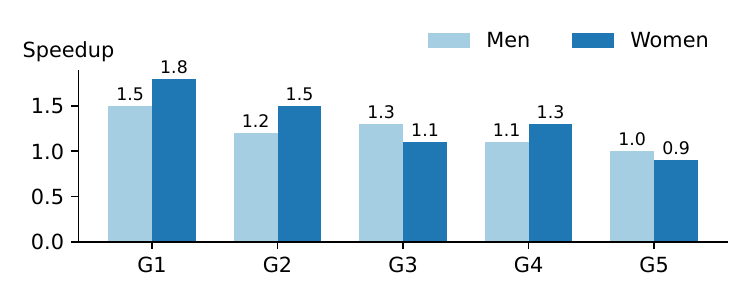}
    \caption{Improved running speed after 4 weeks of training.
    }
    \label{fig:speedup}
  \end{figure}

  \subsubsection{Tables} We optimize our tables for readability by removing as
  much clutter as possible, while highlighting the key structure. Markus Püschel
  (see doc/paper-writing/guide-tables.pdf) wrote a nice guide on how to make nice
  tables. \autoref{tab:simple_table} illustrates this with a simple
  table.

  \begin{table}
    \ra{1.2}
    \centering
    \begin{tabular}{l l l r}
      \toprule
      \textbf{Animal} & \textbf{Size} & \textbf{Biotope} & \textbf{Age} \\
      \midrule
      Dog             & Medium        & Ground           & 20           \\
      Cat             & Medium        & Ground           & 20           \\
      Ant             & Small         & Ground           & 30           \\
      Elephant        & Large         & Ground           & 70           \\
      Whale           & Large         & Water            & 100          \\
      Salmon          & Medium        & Water            & 13           \\
      Eagle           & Large         & Air              & 35           \\
      \bottomrule
    \end{tabular}
    \vspace{1em}
    \caption{A table with heigh lines and emphasized header.}
    \label{tab:simple_table}
  \end{table}

  \subsubsection{Listings} We aim to use minted to create listings as much as
  possible, as this allows us to edit code quickly. We use syntax highlighting
  to make the parts of the code that matter most stand out. Hence, we keep
  most code black, comments gray, and highlight just the MLIR operands that
  we care about most.

  \begin{listing}[H]
    \begin{mlir}
      // This is a comment
      def @foo(
      {
        }
    \end{mlir}
    \caption{A simple MLIR code example with markers. Markers can also be placed in
      captions and refer to labels, e.g. \circled[lst:example]{a}.}
    \label{lst:example}
  \end{listing}

  \begin{listing}[H]
    \begin{lean4}
      theorem funext {f₁ f₂ : ∀ (x : α), β x}
      (h : ∀ x, f₁ x = f₂ x) : f₁ = f₂ := by
      show extfunApp (Quotient.mk' f₁) =
      extfunApp (Quotient.mk' f₂)
      apply congrArg
      apply Quotient.sound
      exact h
    \end{lean4}
    \caption{A simple Lean4 code example, taken from
      \url{https://lean-lang.org/lean4/doc/syntax\_highlight\_in\_latex.html\#example-with-minted}.}
  \end{listing}

  The syntax highlighting also works for xDSL-like IRs.
  Notice that different minted styles can be used for different environments.
  The xDSL environment uses the murphy-style in this case, whereas the MLIR version applies the colorful-style.

  \begin{xdsl*}{fontsize=\scriptsize}
    func.func() [sym_name = "main", function_type = !fun<[
    !iterators.columnar_batch<!tuple<[!i64]>>
    ], []>] {
    ^bb0(
    iterators.scan_columnar_batch(
    iterators.filter(
    iterators.sink(
    func.return()
    }

    func.func() [sym_name = "s0", function_type = !fun<[
    !llvm.struct<[!i64]>], [!i1]>] {
    ^bb0(
    [position = [0 : !index]]
      [predicate = 4 : !i64]
    func.return(
    }
  \end{xdsl*}

  The code in this document was compiled with minted version: \csname ver@minted.sty\endcsname.

  \section{Writing}

  A couple of hints with respect to how we write text.

  \subsection{Citations}
  \label{appendix:citations}

  \subsubsection{Do not use numerical citations as nouns}
  Especially when working with numerical citations (e.g., [1]) the use of
  citations as nouns reduces readability. Hence, we do not use numerical citations
  as nouns and instead expand these citations with \texttt{\textbackslash{}citet} to the
  authornames.
  \\
  {\color{pairedTwoDarkBlue}
  \textit{[1] showed that .. $\dots$}}\\
  $\to$ {\color{pairedFourDarkGreen}\textit{Author et al. [1] showed}}

  \subsubsection{Prefer meaningful text over citations as textual content}
  While acknowledging authors of work is important, maximizing the amount of technical
  content (outside of a historic perspective) typically makes text more direct and
  concrete. Hence, we avoid the discussion of who did what in text if the
  historic context does not add meaning or empty words can be replaced by an immediate
  citation. E.g, in the following the words `introduced in` are not carrying
  information and can be dropped.\\
  {\color{pairedTwoDarkBlue}
  \textit{we extend PreviousIdea introduced in [1] $\dots$ by}}\\
  $\to$ {\color{pairedFourDarkGreen}\textit{we extend PreviousIdea [1] by}}

  \subsubsection{Managing acronyms automatically}
  Managing acronyms manually can lead to situations where the specific term is not properly expanded upon first use or when it is introduced.
  The \texttt{acronym} package is useful to avoid such situations and provides full control over acronyms.
  The expanded form of an abbreviation should be in lowercase, unless its parts are also capitalized (e.g., United Kingdom for UK).
  For example, assume we have defined an acronym with \texttt{\textbackslash{}newacronym\{ir\}\{IR\}\{intermediate representation\}}:
  \begin{itemize}
    \item Upon first use of \texttt{\textbackslash{}ac\{ir\}} we get: \ac{ir}.
    \item On the second reference: \ac{ir}.
    \item To force expansion (e.g., for the background section where the term is first described), we use \texttt{\textbackslash{}acf\{ir\}} which gives: \acf{ir}.
    \item To force contraction (e.g., to save space for a figure caption), we use \texttt{\textbackslash{}acs\{ir\}} which gives: \acs{ir}.
    \item To obtain plural form, we use \texttt{\textbackslash{}acp\{ir\}} giving: \acp{ir}.
  \end{itemize}

  \subsubsection{Adding hyphenation rules}
  While \LaTeX\ handles word breaks automatically, and packages like \texttt{microtype} aim to minimize word splitting, there are instances where either new words lack hyphenation rules, or the suggested hyphenation for a word is undesirable.
  The \texttt{hyphenat} package allows adding hyphenation rules using the \texttt{\textbackslash{}hyphenation} macro, e.g., \texttt{\textbackslash{}hyphenation\{Alex-Net\}} for AlexNet.

  Allowing hyphenation of compound words, we can use \texttt{\textbackslash{}-/} from the \texttt{extdash} package, for example \texttt{high\-/level} can be written as \texttt{high\textbackslash{}-/level}.
  Disallowing a line break at the compound word hyphen, we can use \texttt{\textbackslash{}=/}, as \texttt{RISC\textbackslash{}=/V} for \texttt{RISC\=/V}.

  \section{Example of space saving figure}
See \autoref{fig:sample_space_saving_figure}

\begin{figure}[h]
  \begin{overpic}[width=0.5\textwidth]{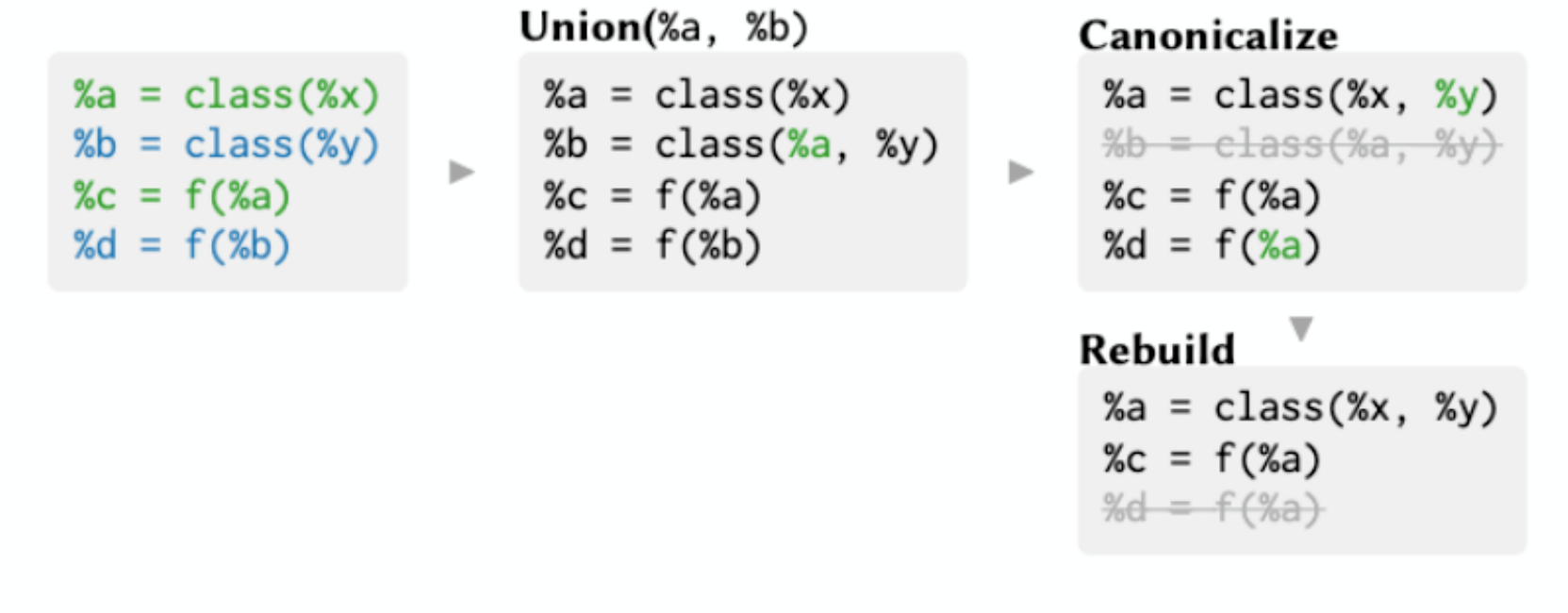}
    \put(2, 11){
      \begin{minipage}{0.30\textwidth}
        \captionof{figure}{Your caption text goes here, sitting inside
        the bottom-left whitespace of the image.}\label{fig:sample_space_saving_figure}
      \end{minipage}
    }
  \end{overpic}
\end{figure}

\end{draftonly}

\end{document}